\documentclass[12pt,a4paper,dvips,twoside,titlepage,fleqn]{article}
\usepackage{calc,xspace}
\usepackage{cite,mcite}
\usepackage{rotating,graphicx}
\usepackage{a4p}
\usepackage{physics}
\usepackage{l3_title,ifthen,Lep}

\graphicspath{{./}}
\DeclareGraphicsExtensions{.eps,.ps,.eps.gz,.ps.gz}
\journalname{Phys. Lett. B}
\date{December 17, 1999}
\preprint{99-181}
\Lep{2}

\newlength{\capindent}
\newlength{\capwidth}
\newlength{\figindent}
\newlength{\figwidth}
\newcommand{\icaption}[2][!*!,!]%
{\hspace*{\capindent}
  \begin{minipage}{\capwidth}
    \ifthenelse{\equal{#1}{!*!,!}}%
    {\caption{#2}}%
    {\caption[#1]{#2}}
  \end{minipage}}

\newcommand{\CoM}{centre-of-mass}
\newcommand{\SM}{Standard Model}
\newcommand{\sqrtsp}{\ensuremath{\sqrt{s^{\prime}}}}
\newcommand{\mumu}{\ensuremath{\mu^{+}\mu^{-}}}
\newcommand{\ff}{\ensuremath{f\kern 0.15em\overline{\kern -0.25em f}}}
\newcommand{\Nsel}{\ensuremath{N_\mathrm{sel}}}
\newcommand{\Nf}{\ensuremath{N_{\mathrm{f}}}}
\newcommand{\Nb}{\ensuremath{N_{\mathrm{b}}}}
\newcommand{\plmi}[2]{^{+\, #1}_{-\, #2}}

\newcommand{\AfbSM}{\ensuremath{A_{\mathrm{fb}}^{\mathrm{SM}}}}
\newcommand{\sigsm}{\ensuremath{\sigma_{\mathrm{SM}}}}

\newcommand{\cost}{\ensuremath{\cos\theta}}
\newcommand{\minv}{\ensuremath{m_{\ffbar}}}
\newcommand{\epsdif}{\ensuremath{\epsilon(\cost,\minv)}}
\newcommand{\dtsig}[1]{\ensuremath{\frac{\mathrm{d^{2}}\sigma_{#1}}%
    {\mathrm{d}\!\cost\,\mathrm{d}\minv}}}

\newcommand{\phm}{\phantom{-}}

\begin{document}
       
\begin{titlepage}
  \title{\boldmath
    Measurement of Hadron and Lepton-Pair Production \\
    at $130 \GeV < \sqrt{s} < 189 \GeV$ at LEP}
  \author{{\Large L3 Collaboration}}
  \begin{abstract}
    We report on measurements of {\epem} annihilation into hadrons and lepton
    pairs. The data have been collected with the L3 detector at LEP at {\CoM}
    energies between 130 and 189 \GeV.
    Using a total integrated luminosity of 243.7~\pb, 25864
    hadronic and 8573 lepton-pair events are selected for the
    measurement of cross sections and leptonic forward-backward asymmetries. The
    results are in good agreement with {\SM} predictions.
  \end{abstract}
\submitted
\end{titlepage}

\section{Introduction}

We report on the results of measurements of fermion-pair production above the Z
pole, based on data collected using the L3 detector at LEP in 1997 and 1998 at
\CoM\ energies $\sqrt{s} = 182.7 \GeV$ and $\sqrt{s} = 188.7 \GeV$,
respectively. Data corresponding to integrated luminosities of 55.5~\pb\ and
176.2~\pb\ were collected, leading to much improved statistics compared to our
previous publications~\cite{l3-90,l3-117} based on data from 1995 and
1996. In addition, in 1997, small amounts of data, 3.4~\pb\ and 3.6~\pb, were
collected at the same \CoM\ energies as in 1995, 130.0~\GeV\ and 136.1~\GeV,
respectively. The measurements made on these data samples are combined with those
resulting from a re-analysis of the previous data, superseding the
results obtained in Reference~\cite{l3-90}.

In this article we report on measurements of the fermion pair production
reactions:
\begin{eqnarray*}
 \epem\rightarrow\mbox{hadrons}(\gamma)\, , ~~~~
 \epem\rightarrow\mumu(\gamma)\, , ~~~~
 \epem\rightarrow\tautau(\gamma)\, , ~~~~
 \epem\rightarrow\epem(\gamma)\, . &&
\end{eqnarray*}
In these reactions, the $(\gamma)$ indicates the possible presence of additional
photons or low invariant-mass fermion pairs.

For a substantial fraction of the events initial-state radiation, ISR,
lowers the initial \CoM\ energy to an \emph{effective} \CoM\ energy of the
annihilation process, \sqrtsp.
When \sqrtsp\ is close to the Z mass, \MZ, the events are classified as radiative
returns to the Z.
A cut on \sqrtsp\ allows a separation between events at high effective \CoM\
energies (so-called \emph{high-energy} events), and radiative returns to the Z.
Cross sections are measured for all processes and forward-backward asymmetries
are measured for the lepton channels and are compared to predictions of the
\SM~\cite{SM,SM-2}, both for the high-energy sample and for a larger,
\emph{inclusive} sample including also the radiative returns to the Z.
Kinematic cuts have been changed with respect to our previous
publication~\cite{l3-117}. The corresponding results of the cross section and
forward-backward asymmetry measurements have been included, with corrections for
these changes applied.

Similar studies on the data taken at centre-of-mass energies between 182.7~\GeV\
and 188.7~\GeV\ have been published by other
LEP collaborations~\cite{opal-245}.

\section{Analysis Method}
\label{sec:analysis}

The data were collected using the L3 detector described
in References~\cite{l3-00,L3-SLUM}.
For the $s$-channel processes, the inclusive event sample is defined by requiring
$\sqrtsp > 60 \GeV$ for hadronic events and $\sqrtsp > 75 \GeV$ for lepton-pair
events, to reduce uncertainties on radiative corrections in extrapolating to low
\sqrtsp\ values. The high-energy sample is defined by requiring $\sqrtsp > 0.85
\sqrt{s}$.

Using the sum of all ISR photon or pair energies, $E_{\gamma}$, and momentum
vectors, ${\mathbf{P}}_{\gamma}$, the $s^{\prime}$ value is given by:
\begin{equation} \label{eq:spri}
  s^{\prime} \, = \, s - 2 E_\gamma\sqrt{s} + E_{\gamma}^2 - {\bf P}_{\gamma}^2 \, .
\end{equation}

For most of the events initial-state radiation is along the beam pipe and is
not detected. In this case a single photon is assumed to be emitted along the
beam axis; its energy is determined from the event kinematics.
The \sqrtsp\ value is estimated using Equation~\ref{eq:spri}. The effect of
multiple photon and final-state radiation on the \sqrtsp\ calculation has been
studied using Monte Carlo programs and is corrected for. The treatment of photons
observed in the detector is addressed in the sections describing the individual
analyses. Mis-reconstruction of the effective \CoM\ energy induces a migration of
events between the kinematic regions allowed and excluded by the cut on
\sqrtsp. This is taken into account in the efficiency determination and as an
additional background, denominated as ISR contamination.

Bhabha scattering at high energies is dominated by $t$-channel photon exchange,
and hence a cut on $s^{\prime}$ is less natural. Instead, a cut is applied on the
acollinearity angle, $\zeta$, of the final-state $\e^{+}$ and $\e^{-}$. In this
case, the inclusive and high-energy samples are defined by requiring $\zeta <
120^{\circ}$ and $\zeta < 25^{\circ}$, respectively.

Selection efficiencies and backgrounds are determined by Monte Carlo
simulations for each {\CoM} energy using the following event
generators: BHLUMI~\cite{BHLUMI} (small-angle Bhabha scattering);
PYTHIA~\cite{PYTHIA} ($\epem\rightarrow\mbox{hadrons}(\gamma)$,
$\mathrm{ZZ}(\gamma)$, $\mathrm{Z}\e\e(\gamma)$, $\mathrm{W}\e\nu(\gamma)$);
KORALZ~\cite{KORALZ} ($\epem\rightarrow\mumu(\gamma)$,
$\epem\rightarrow\tautau(\gamma)$);
BHAGENE~\cite{BHAGENE} and BHWIDE~\cite{BHWIDE} (large-angle Bhabha scattering);
TEEGG~\cite{TEEGG} ($\epem\rightarrow\epem\gamma(\gamma)$);
GGG~\cite{GGG} ($\epem\rightarrow\gamma\gamma(\gamma)$);
PHOJET~\cite{PHOJET} (hadronic two-photon collisions);
DIAG36~\cite{DIAG} ($\epem\rightarrow\epem\mumu$,
$\epem\tautau$, $\epem\epem$);
FERMISV~\cite{FERMIS} ($\epem\rightarrow\epem\ffbar$);
KORALW~\cite{KORALW} ($\epem\rightarrow\mathrm{W^{+}W^{-}}(\gamma)$);
and EXCALIBUR~\cite{EXCALIBUR}
($\epem\rightarrow\mathrm{q\bar{q^{\prime}}}\e\nu(\gamma)$,
$\epem\rightarrow\epem\epem$).

The measurements are compared to the predictions of the \SM\ as calculated using the
ZFITTER~\cite{ZFITTER} and TOPAZ0~\cite{TOPAZ0} programs with the following
parameters~\cite{l3-69,l3-55,PDG,JEGERLEHNER,cdf-mtop-1}: 
$\MZ=91.190$~{\GeV},
$\alpha_{s}(\MZ^2)=0.119$,
$\Mt=173.8$~{\GeV},
$\Delta\alpha_{\mathrm{had}}^{(5)}=0.02804$,
and $\MH=150$~{\GeV}. The theoretical uncertainties on the \SM\ predictions are
estimated to be below 1\%~\cite{err-zf} except for the
predictions for large angle Bhabha scattering which have an uncertainty of
2\%~\cite{BHABHA-THEORY}.

\subsection{Initial-final state interference in \boldmath $s$-channel processes}
\label{sec:ifsr}

In the presence of interference between initial- and final-state radiative
corrections, the effective \CoM\ energy, in contrast to the acollinearity angle,
is not well-defined.
Moreover, for the $s$-channel processes, unlike for Bhabha scattering, these
contributions are not included in the Monte Carlo samples used to estimate
efficiencies.
Their effect is expected to be largest for the high-energy
\mumu\ and \tautau\ samples, affecting cross sections by up to 2\% and
forward-backward asymmetries by up to 0.02. The following approach is used in
the analysis of the $s$-channel processes.

Cross sections are first determined disregarding this effect. This allows the use
of existing Monte Carlo programs without modifications. Corrections are
subsequently applied using the \SM\ predictions for the interference
contributions, folded with the selection efficiency, $\epsilon$, as a function of
the fermion-pair invariant mass, \minv, and the scattering angle of the
anti-fermion, \cost. This leads to an additive correction:
\begin{equation}
  \label{eq:ifsr-cross}
  \sigma_{\ffbar} \rightarrow \sigma_{\ffbar} -
  \frac{1}{\epsilon}\int_{-1}^{1}\mathrm{d}\!\cost 
  \int_{\sqrtsp}^{\sqrt{s}}\,\mathrm{d}\minv\,\epsdif\,\dtsig{\mathrm{intf}},
\end{equation}
where $\mathrm{d^{2}}\sigma_{\mathrm{intf}}/\mathrm{d}\!\cost\,\mathrm{d}\minv $
is the differential interference contribution to the cross section as calculated
using the ZFITTER program. The corrections applied range between 0.1\% for the
hadronic cross section and 1.3\% for the leptonic cross sections.

The forward-backward asymmetries for the high-energy samples are obtained as
the result of an unbinned maximum-likelihood fit to the polar angular
distribution in the Born approximation, to which is added a term representing the
differential interference cross section:
\begin{equation}
  \label{eq:cost-afbfit}
  \frac{1}{\sigma_{\ffbar}}\frac{\mathrm{d}\sigma_{\ffbar}}{\mathrm{d}\!\cost} =
  \frac{3}{8}(1+\cos^{2}\theta) + \Afb \cost +
  \frac{1}{\epsilon(\cost)\sigma_{\ffbar}}
  \int_{\sqrtsp}^{\sqrt{s}}\mathrm{d}\minv\,\epsdif\,\dtsig{\mathrm{intf}} \: .
\end{equation}
For the inclusive sample, initial-state radiation distorts the angular
distribution, such that the Born approximation in Equation~\ref{eq:cost-afbfit} is
not appropriate. Instead, the forward-backward asymmetry is obtained directly
from the differential cross section and extrapolated to the full solid angle using
the ZFITTER program. The
differential cross section is corrected analogously to
Equation~\ref{eq:ifsr-cross}. The correction is largest for the asymmetry of the
high-energy samples, ranging between 0.004 and 0.010.

\subsection{Pair corrections}
\label{sec:ispp}

Besides the emission of ISR photons, also the emission of initial-state pairs
can lower the effective centre-of-mass energy of the scattering process.
This gives rise to a non-negligible contribution to the inclusive cross section
(approximately 1.5\% for all $s$-channels as estimated using the ZFITTER
program) when
radiative returns to the Z are included in the signal definition. To allow for a
proper comparison between experimental measurements and theoretical predictions,
these radiative corrections are included in the fermion-pair signal definition.

To calculate the effect of this signal contribution on the overall efficiency, and
to estimate the background contributions leading to the same four-fermion final
states, events are generated using the DIAG36 program. As this program includes
only photon exchange, the events are reweighted to include the effects of Z
exchange using the matrix element calculation of the FERMISV program.
The selection efficiencies are obtained by combining those estimated from the
separate Monte Carlo samples regarded as signal, weighted with their
respective cross sections as estimated using the ZFITTER program.
As these Monte Carlo programs do not yield a correct description of
low-mass hadronic pairs, the efficiency for events with hadronic pairs is taken to
be that for the events with lepton pairs. A 20\% uncertainty is assigned to this
efficiency, resulting in an uncertainty less than 0.2\% on the overall efficiency.

Because of the large number of diagrams involved, this approach is less
straightforward in the case of Bhabha scattering. Since the relative pair
correction is estimated~\cite{pairs-ee} to be significantly smaller than for the
$s$-channel processes, its effect on the selection efficiency is neglected and no
correction is applied.

\section{Analysis and Results}
\label{sec:anal}

\subsection{Integrated luminosity}

The luminosity is measured using small-angle Bhabha scattering~\cite{L3-SLUM}. A
tight fiducial volume cut, $34 \mbox{ mrad} < \theta < 54 \mbox{ mrad}$ and
$|90^{\circ}-\phi| > 11.25^{\circ}$, $|270^{\circ}-\phi| > 11.25^{\circ}$, is
imposed on the coordinates of the highest-energy cluster on one side. The
highest-energy cluster on the opposite side should be contained in a looser
fiducial volume, $32 \mbox{ mrad} < \theta < 65 \mbox{ mrad}$ and 
$|90^{\circ}-\phi| > 3.75^{\circ}$, $|270^{\circ}-\phi| > 3.75^{\circ}$. This
method reduces the theoretical uncertainty.

The experimental systematic uncertainties originate from the event selection
criteria, 0.10\%, and from the detector geometry, 0.05\%.
The Monte Carlo statistics result in an uncertainty of 0.07\%,
yielding a total experimental systematic uncertainty of 0.13\%.
In addition, a theoretical uncertainty of 0.12\%~\cite{BHLUMI-new} is assigned
to the BHLUMI generator, resulting in a total uncertainty of 0.18\%.

\subsection{\boldmath $\epem\rightarrow\mbox{hadrons}(\gamma)$}
 
\subsubsection*{Event selection}

Events are selected by restricting the visible energy, 
$ E_{\mathrm{vis}} $, to $ 0.4 < E_{\mathrm{vis}}/\sqrt{s} < 2.0 $.
The longitudinal energy imbalance must satisfy
$|E_{\mathrm{long}}|/E_{\mathrm{vis}}<0.7$. The reconstructed energies do not
include isolated electromagnetic energy depositions with an energy greater than
10 \GeV. These cuts reject most of the background from two-photon collision
processes.

In order to reject background originating from lepton pair events, more than 18
calorimetric clusters with an energy exceeding 300 \MeV\ each are requested. 

The W-pair production background is reduced by applying the following cuts.
Semi-leptonic W-pair decays are rejected by requiring the transverse
energy imbalance to be smaller than 0.3 $E_{\mathrm{vis}}$. The background from
hadronic W-pair decays is reduced by rejecting events with at least 
four jets each with energy greater than 15 \GeV. The jets are obtained
using the JADE~\cite{jade} algorithm with a fixed jet resolution parameter 
$y_{\mathrm{cut}}=0.01$.

\begin{figure}[p]
  \begin{center}
    \includegraphics[width=0.48\textwidth]{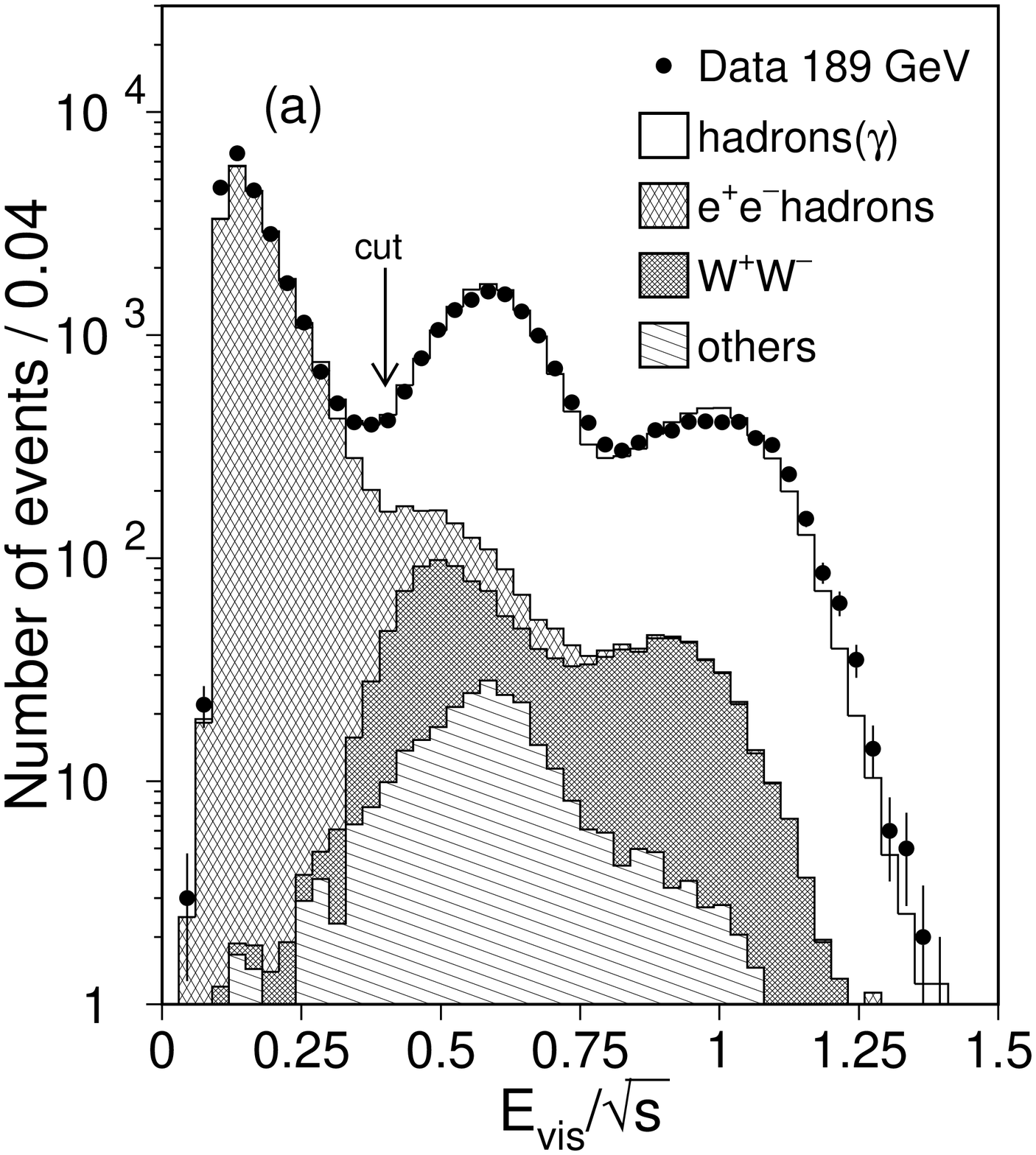}
    \includegraphics[width=0.48\textwidth]{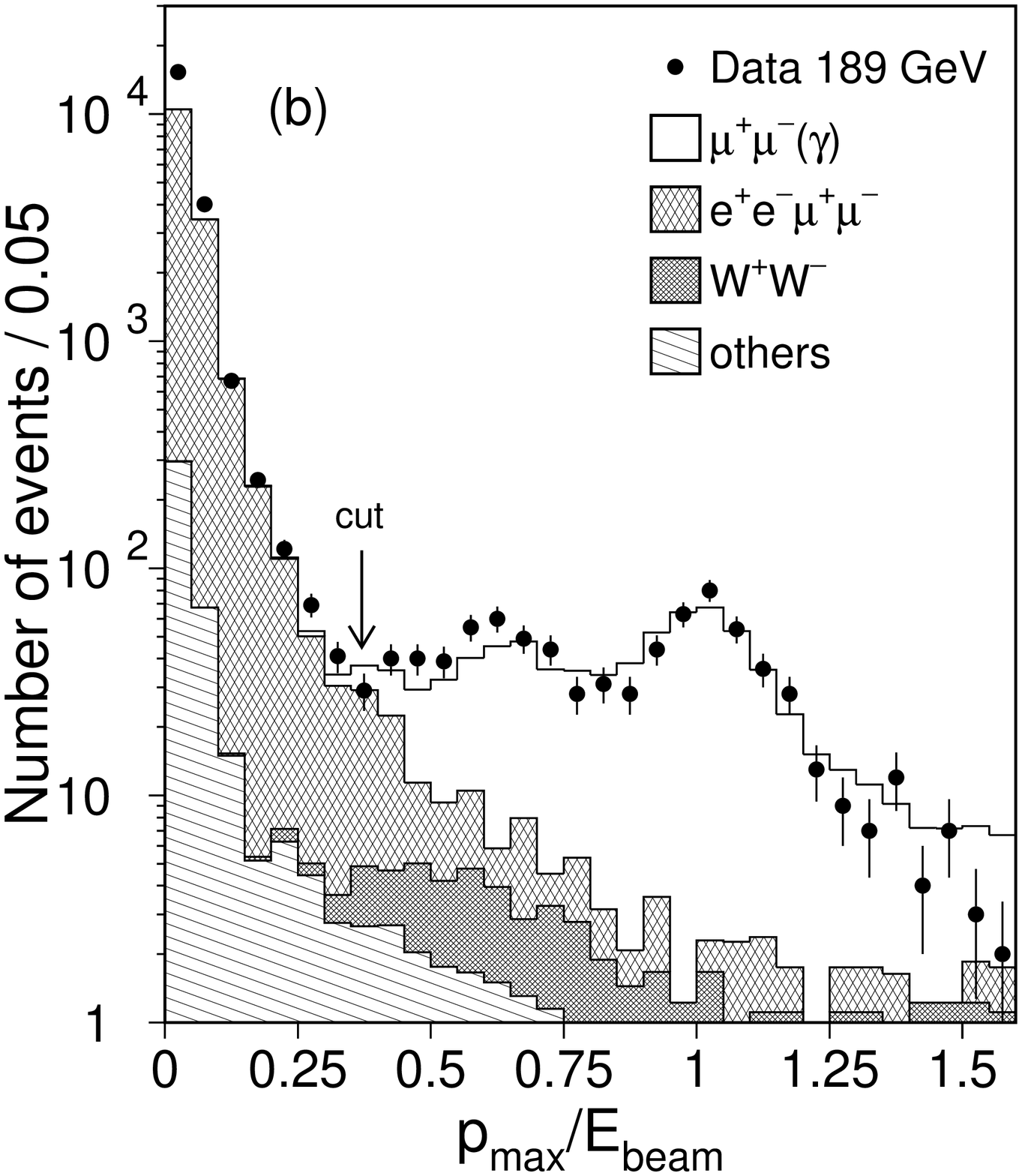}
    \includegraphics[width=0.48\textwidth]{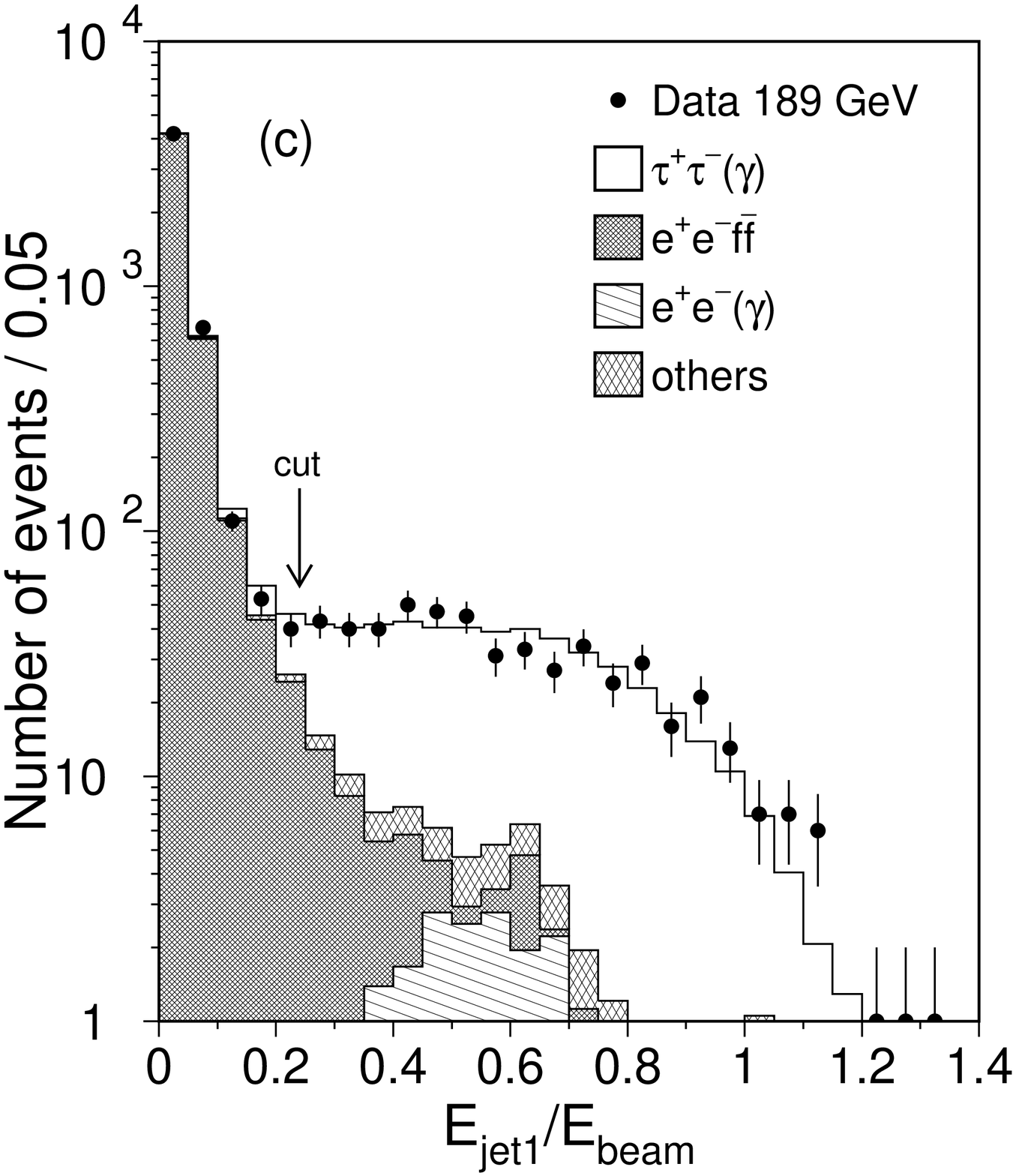}
    \includegraphics[width=0.48\textwidth]{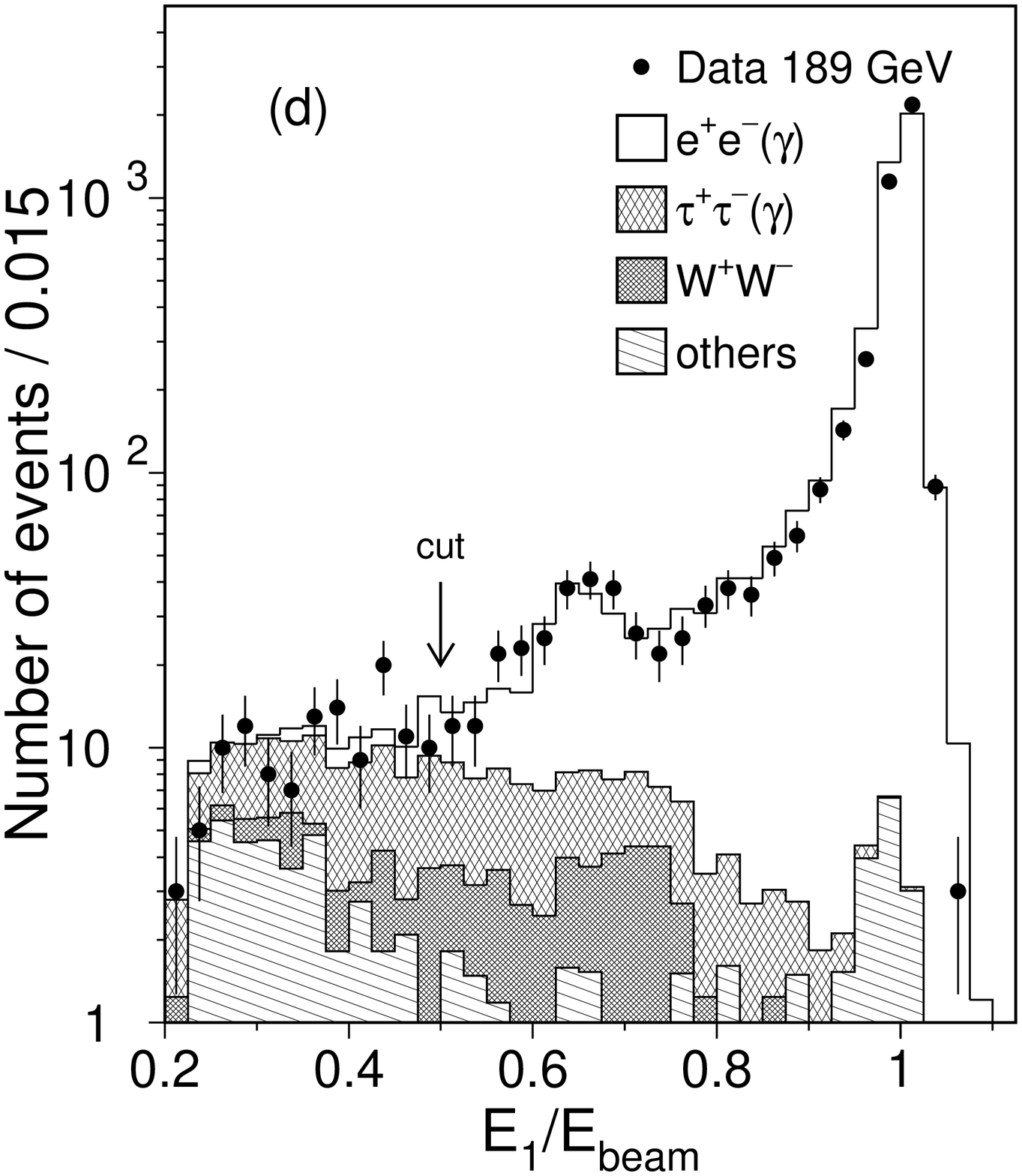}
    \parbox{\capwidth}{
      \caption[]{
        (a) The total visible energy normalised to the {\CoM} energy,
        {$\sqrt{s}$}, for the selection of
        {$\epem\rightarrow\mbox{hadrons}(\gamma)$} events,
        (b) highest muon momentum normalised to the beam energy for the selection
        of {$\epem\rightarrow\mumu(\gamma)$} events,
        (c) highest tau jet energy normalised to the beam energy for the
        selection of {$\epem\rightarrow\tautau(\gamma)$} events, and
        (d) highest electron energy normalised to the beam energy for
        the selection of {$\epem\rightarrow\epem(\gamma)$} events.}
      \label{fig:sele}}
  \end{center}
\end{figure}

Figure~\ref{fig:sele}a shows the distribution of the visible energy normalised to
the centre-of-mass energy for hadronic final state events selected
at 189 \GeV. The observed peak structure of the signal arises
from the high-energy events and from the radiative returns to the Z.

As an additional cross-check, an alternative selection is performed using an
artificial neural network technique~\cite{Thesis-qq} instead of the cuts
described above. The results obtained using the two selection methods are
compatible with each other.

To reconstruct the effective centre-of-mass energy, two different methods
are used. In the first method, all events are reclustered into two
jets using the JADE algorithm.
A single photon is assumed to be emitted along the beam axis and to result in
a missing momentum vector. From the polar angles of the jets, $\theta_{1}$ and
$\theta_{2}$, the photon energy is then estimated as:
\begin{equation} \label{eq:egam}
E_{\gamma} ~ = ~ \sqrt{s} \cdot \frac{|\sin(\theta_{1}+\theta_{2})|}
           {\sin\theta_{1}+\sin\theta_{2}+|\sin(\theta_{1}+\theta_{2})|} \,.
\end{equation}
The second method uses the clustered jets obtained using the JADE algorithm with a
fixed cut, $y_{\mathrm{cut}}=0.01$.
A kinematic fit is performed assuming the emission of either zero, one, or two
photons along the beam axis. The hypothesis of the smallest number of photons
yielding a probability of the kinematic fit larger than 8.5\% is used.
The cross sections are estimated as the average of the results obtained using the
two methods. A systematic uncertainty on the \sqrtsp\ reconstruction,
equal to half their difference, is assigned.

For about 10\% of the events, a high-energy cluster is detected in the
electromagnetic calorimeter. It is selected as described above and is assumed to
be a photon. Its energy and momentum are added to the undetected ISR photons. 
The effective centre-of-mass energy is then calculated using
Equation~\ref{eq:spri}.

\begin{figure}[p]
  \begin{center}
    \includegraphics[width=0.48\textwidth]{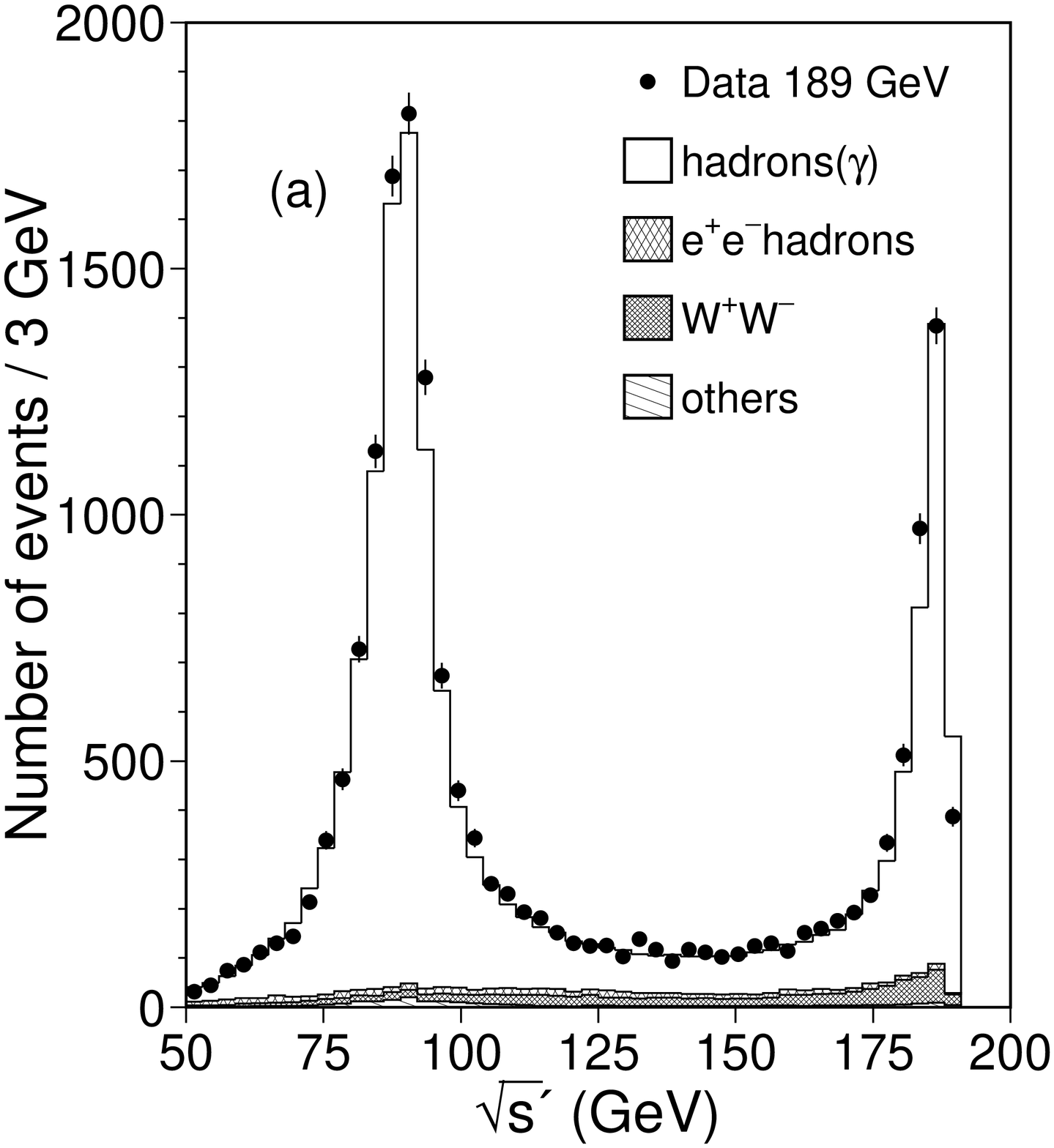}
    \includegraphics[width=0.48\textwidth]{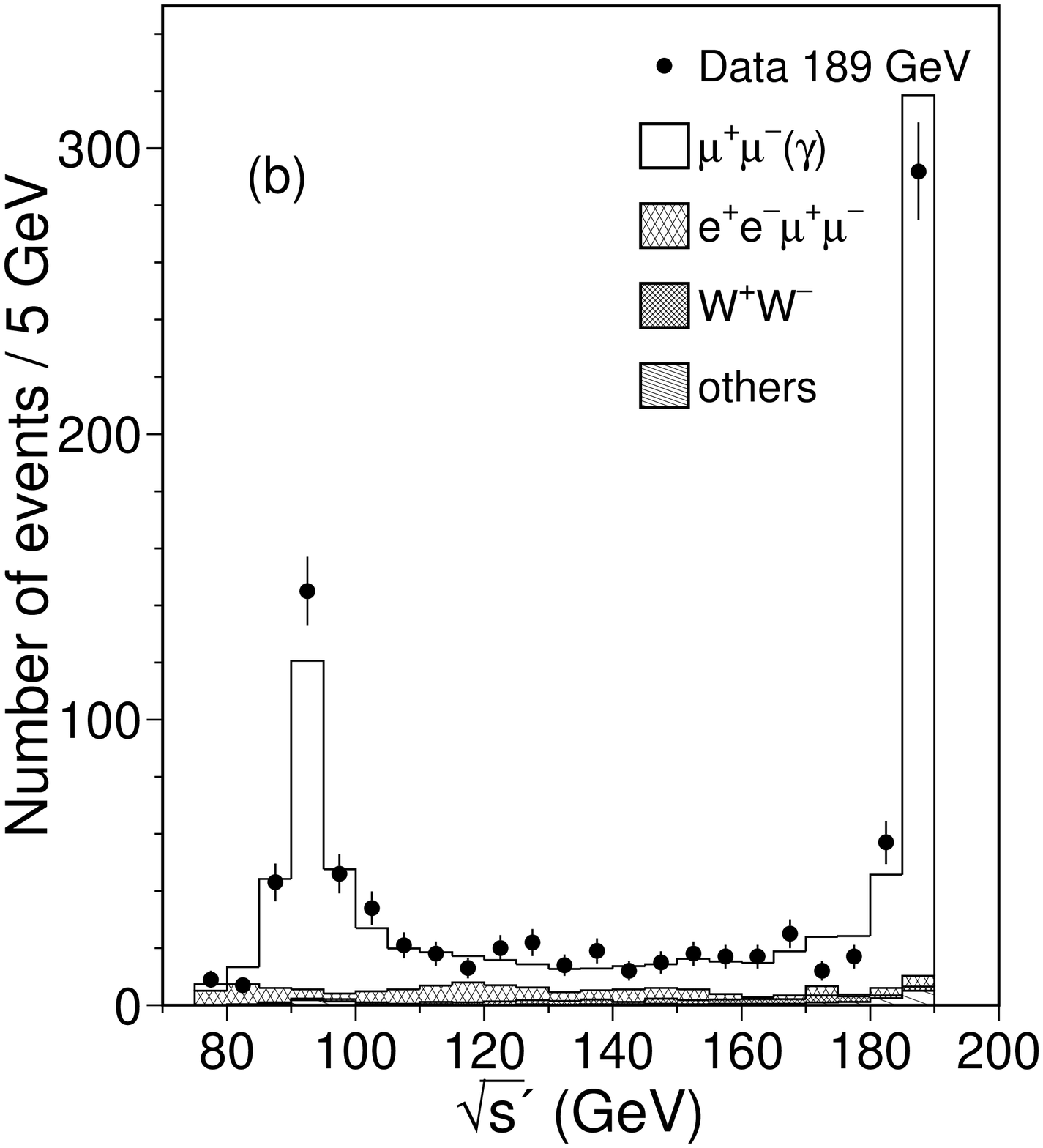}
    \includegraphics[width=0.48\textwidth]{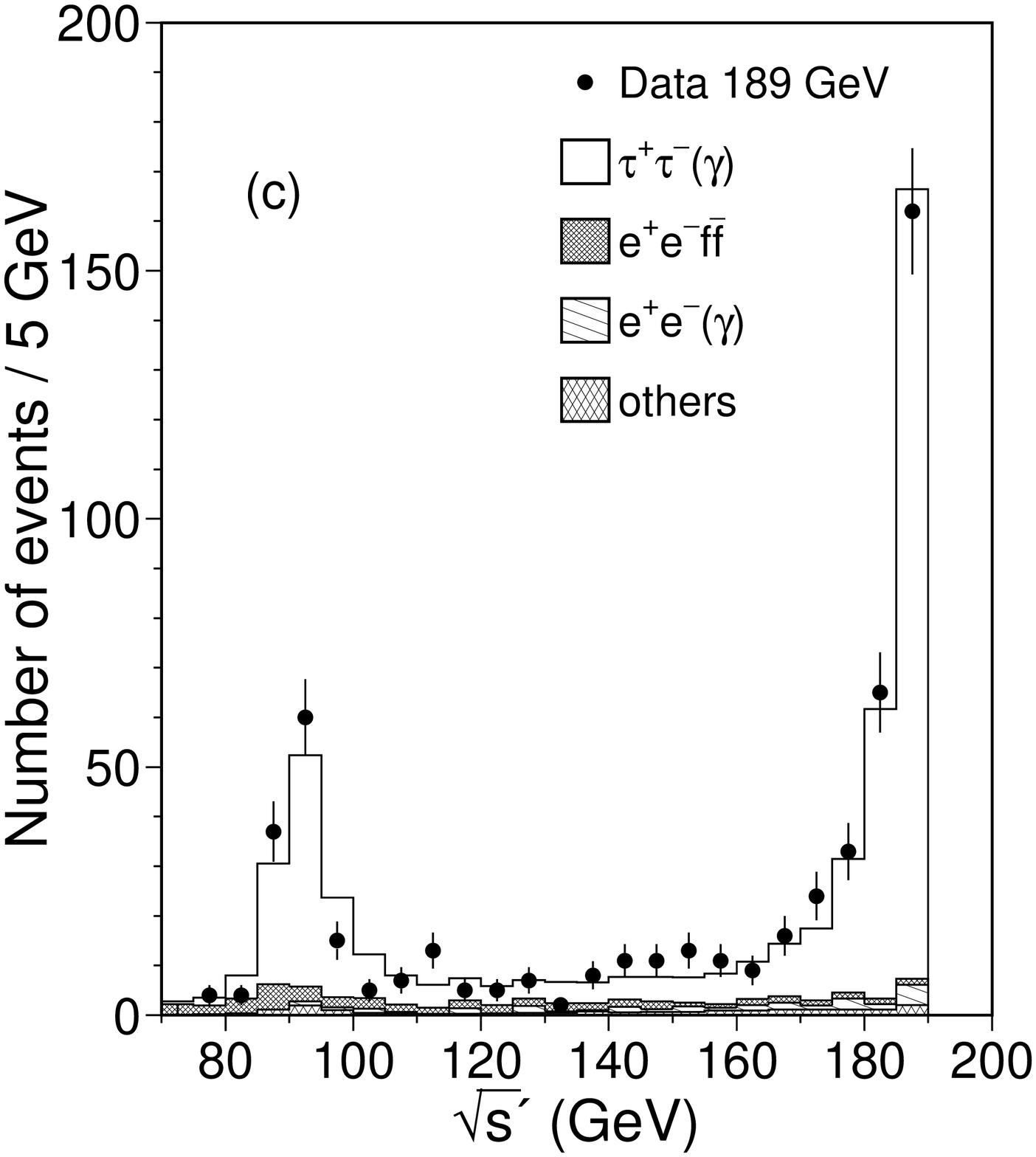}
    \includegraphics[width=0.48\textwidth]{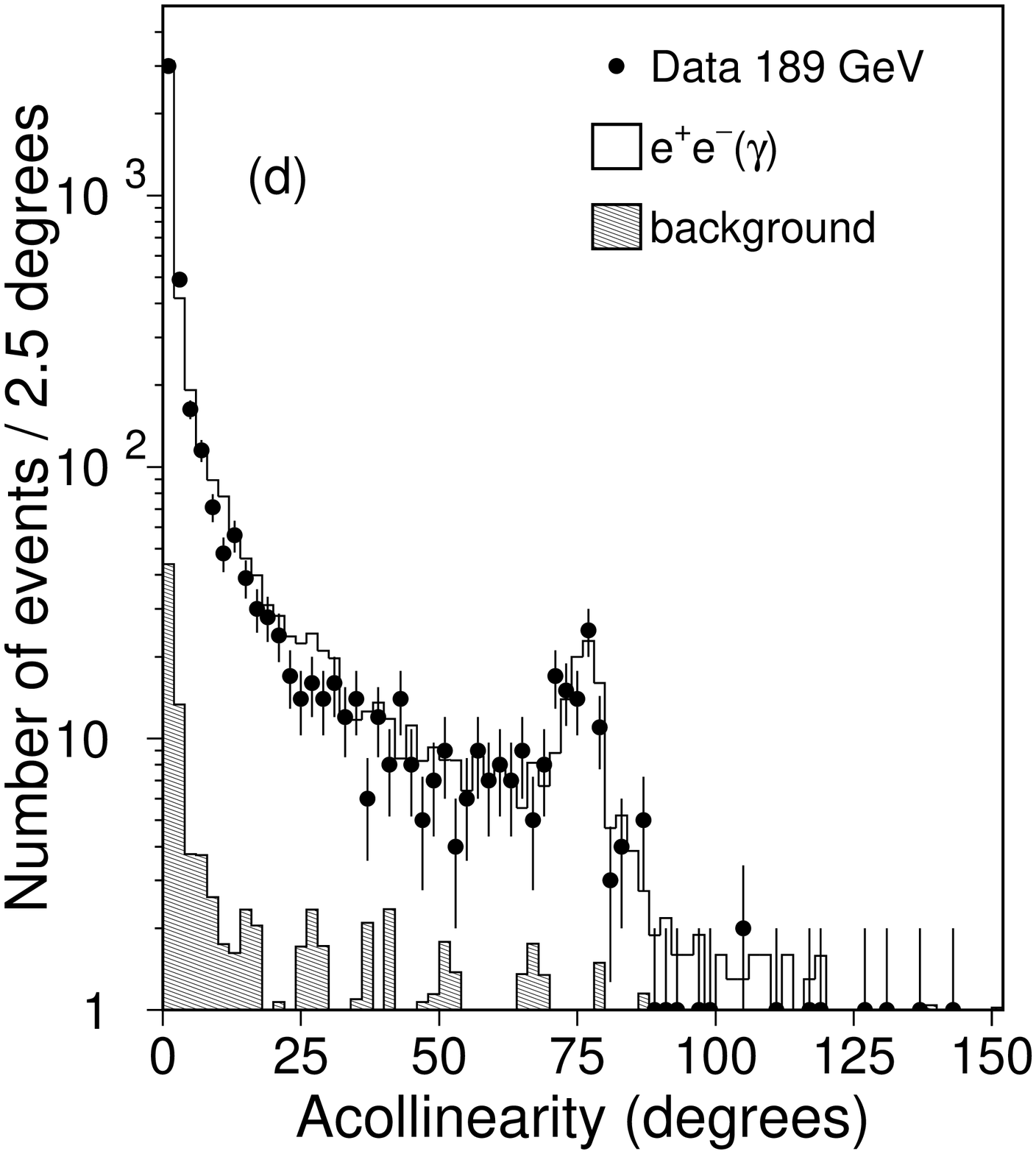}
    \parbox{\capwidth}{
      \caption[]{
        The reconstructed effective {\CoM} energy, {\sqrtsp}, for the selection of
        (a) {$\epem\rightarrow\mbox{hadrons}(\gamma)$} events, (b)
        {$\epem\rightarrow\mumu(\gamma)$} events, (c)
        {$\epem\rightarrow\tautau(\gamma)$} events, and the reconstructed
        acollinearity angle for (d)
        {$\epem\rightarrow\epem(\gamma)$} events.}
      \label{fig:spri}}
  \end{center}
\end{figure}

Figure~\ref{fig:spri}a shows the reconstructed \sqrtsp\ distribution, based on
the reconstruction using the jet angles, for hadronic final state events.

\subsubsection*{Cross section}

Selection efficiencies and background contributions are listed, for the \sqrtsp\
reconstruction method using the jet angles, in Table~\ref{tab:sele}.
The selected sample contains a background from hadronic two-photon collision
processes, W-, Z- and tau-pair production and $\epem\rightarrow
\mathrm{Z}\epem(\gamma)$ events. The two-photon background is estimated by
adjusting the Monte Carlo to the data in a two-photon enriched sample.

\begin{figure}[p]
  \begin{center}
    \includegraphics[width=0.9\textwidth]{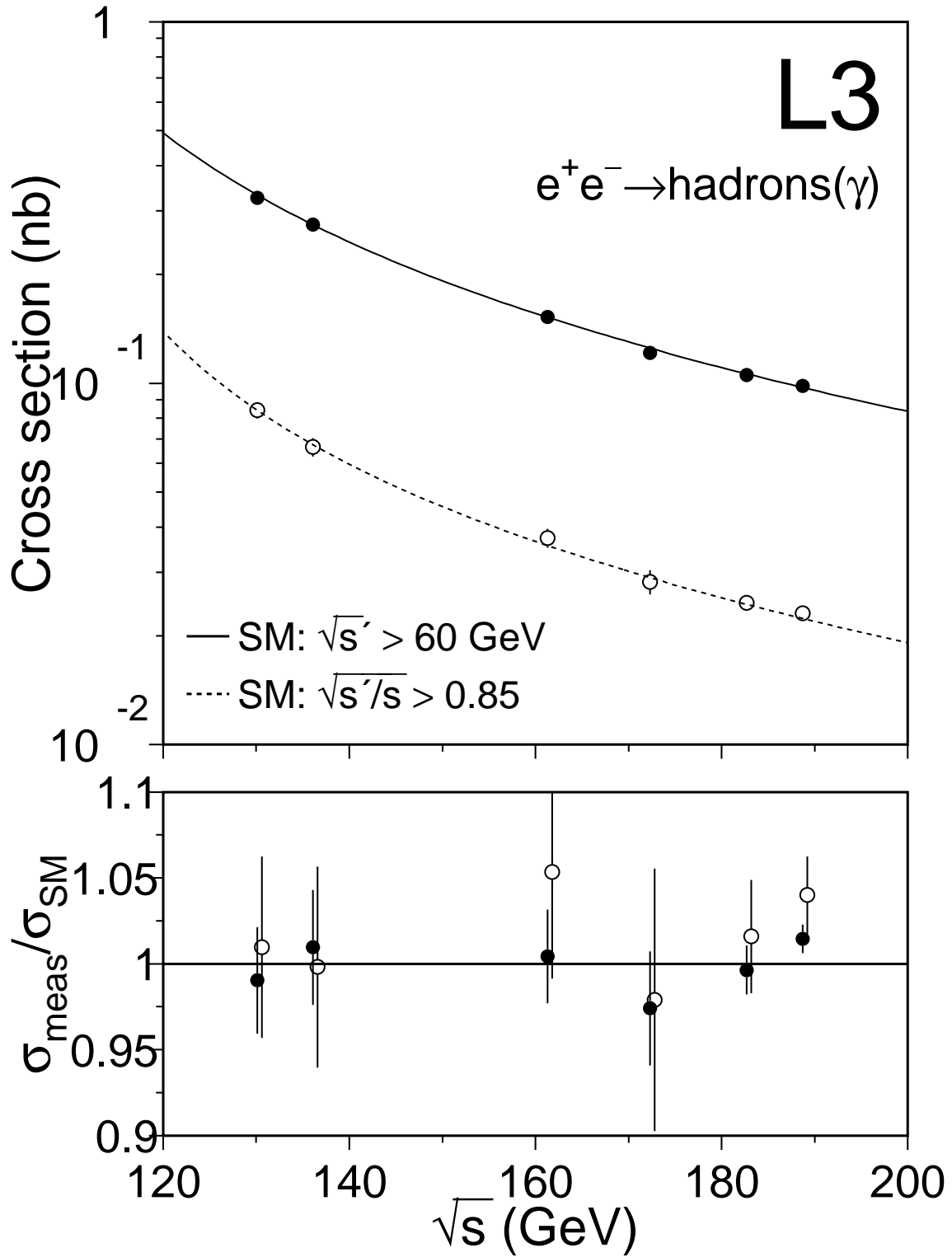}
    \parbox{\capwidth}{
      \caption[]{
        Cross sections of the process {$\epem\rightarrow\mbox{hadrons}(\gamma)$},
        for the inclusive (solid symbols) and the high-energy sample (open symbols).
        The {\SM} predictions are shown as a solid line for the inclusive sample
        and as a dashed line for the high-energy sample. The lower plot shows
        the ratio of measured and predicted cross sections.}
      \label{fig:ha_xsec}}
  \end{center}
\end{figure}

The numbers of selected events, the total cross sections for the different
event samples, and the corresponding statistical and systematic uncertainties
are listed in Table~\ref{tab:xsec}, together with our previous published
measurements~\cite{l3-117}. The systematic uncertainties are dominated
by the uncertainty on the \sqrtsp\ determination and are correlated between
different \CoM\ energies. In Figure~\ref{fig:ha_xsec} the cross section
measurements are shown and compared to the \SM\ predictions.

\subsection{\boldmath $\epem\rightarrow\mumu(\gamma)$}
\label{sec:muon}
\subsubsection*{Event selection}

The event selection for the process {$\epem\rightarrow\mumu(\gamma)$} follows
that of Reference~\cite{l3-117}. Two muons are required within the polar angular
range $|\cos\theta|<0.9$. For the data taken at 183 \GeV\ the angular range is
restricted to $|\cos\theta|<0.81$.
At least one muon must be measured in the muon spectrometer, and have a momentum
greater than 35 \GeV. This reduces substantially the background from
$\epem\rightarrow\epem\mumu$ interactions whilst ensuring a high acceptance for
events with hard ISR photons.

Background from cosmic muons is reduced using both scintillation counter time
information and the distance of the muon tracks from the beam axis. The number of
accepted cosmic muon events 
is estimated by extrapolating the corresponding sideband distributions to the
signal region.
Figure~\ref{fig:sele}b shows the distribution of the maximum muon momentum
normalised to $E_{\mathrm{beam}}$ for events selected at 189~{\GeV}.

The {\sqrtsp} value for each event is determined using Equation~\ref{eq:spri}
assuming the emission of a single ISR photon. In case a photon is detected in the
electromagnetic calorimeter it is required to have an energy greater than 15~{\GeV}
and an angular separation to the nearest muon of more than 10 degrees. Otherwise
the photon is assumed to be emitted along the beam axis and its energy is
calculated from the polar angles of the outgoing muons according to
Equation~\ref{eq:egam}.
The distribution of the reconstructed {\sqrtsp} for events selected at 189~{\GeV}
is shown in Figure~\ref{fig:spri}b.

\subsubsection*{Cross section}

Selection efficiencies and background contributions are listed in
Table~\ref{tab:sele}. The main background contributions are from the reactions
$\epem\rightarrow\epem\mumu$, $\epem\rightarrow\tautau(\gamma)$ and from W-pair
production.

\begin{figure}[p]
  \begin{center}
    \includegraphics[height=0.45\textheight]{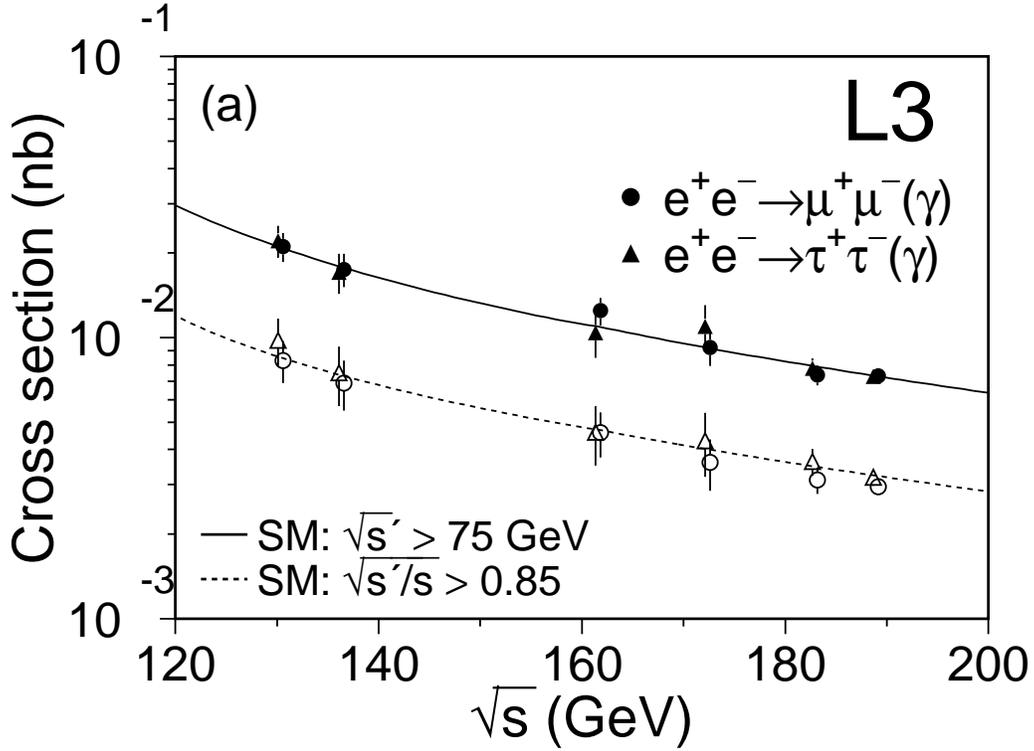}
    \includegraphics[height=0.45\textheight]{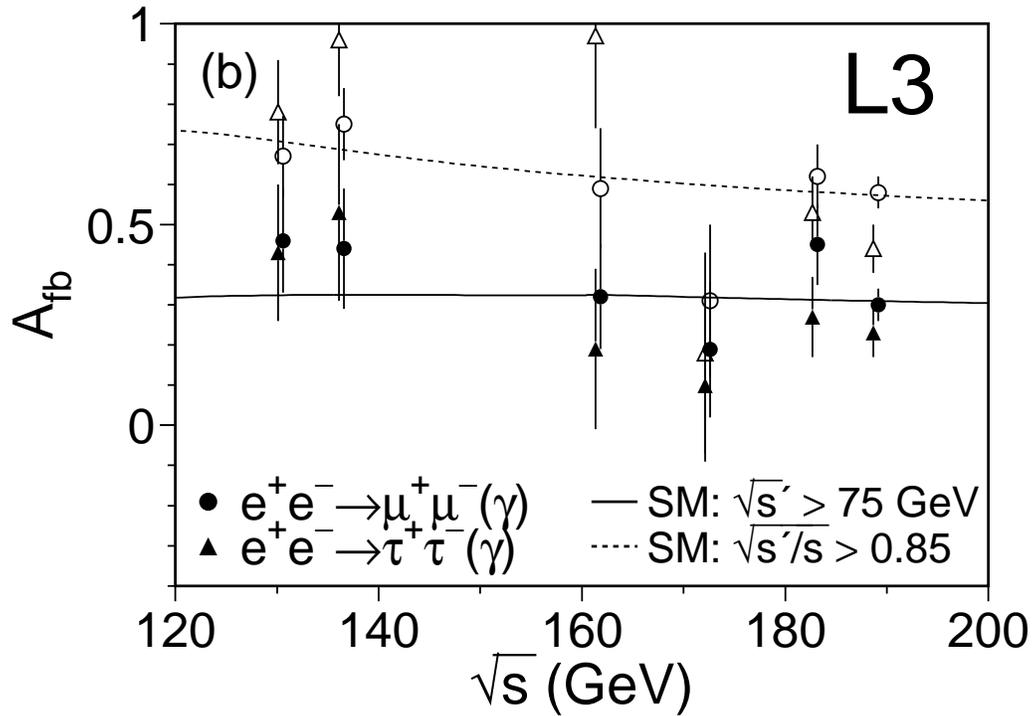}
    \parbox{\capwidth}{
      \caption[]{
        Cross sections (a) and forward-backward asymmetries (b) of the processes
        {$\epem\rightarrow\mumu(\gamma)$} and
        {$\epem\rightarrow\tautau(\gamma)$} for the inclusive (solid symbols)
        and the high-energy sample (open symbols). The {\SM} predictions are shown
        as a solid line for the inclusive sample and as a dashed line for the
        high-energy sample.}
      \label{fig:le_xsec_afb}}
  \end{center}
\end{figure}

Table~\ref{tab:xsec} summarises the numbers of selected events, the resulting
cross sections, and their statistical and systematic uncertainties for the two
event samples at the various {\CoM} energies. The main contributions to the
systematic uncertainties originate from the background subtraction and from the
acceptance correction.
Figure~\ref{fig:le_xsec_afb} shows the comparison to the {\SM} prediction.

\subsubsection*{Forward-backward asymmetry}
\label{sec:mmafb}

The forward-backward asymmetry is determined using events with two
muons with opposite charge and an acollinearity angle smaller than 90~degrees.

For the high-energy sample, the angular distribution of the events is
parametrised according to Equation~\ref{eq:cost-afbfit}.
The asymmetry, \Afb, is determined from an unbinned maximum-likelihood fit 
of this parametrisation to the data within the fiducial volume. The muon
charge is measured as described in Reference~\cite{l3-117}. The charge confusion
per event, ranging between  0.2\% and 0.7\%, is taken into account in the fit
procedure.
The asymmetries for the accepted background contributions are estimated using the
same method and are corrected for. The corrections range between 0.045 and 0.059.

For the inclusive event sample the differential cross section is distorted by hard
ISR photons. Therefore, \Afb\ is computed directly from the differential cross
sections obtained within the fiducial volume. To obtain the asymmetry for the
full solid angle an extrapolation factor is calculated using the ZFITTER program.
It ranges between 1.10 for the 183 \GeV\ data and 1.03 for the 189 \GeV\ data.

\begin{figure}[p]
  \begin{center}
    \includegraphics[width=1.\textwidth]{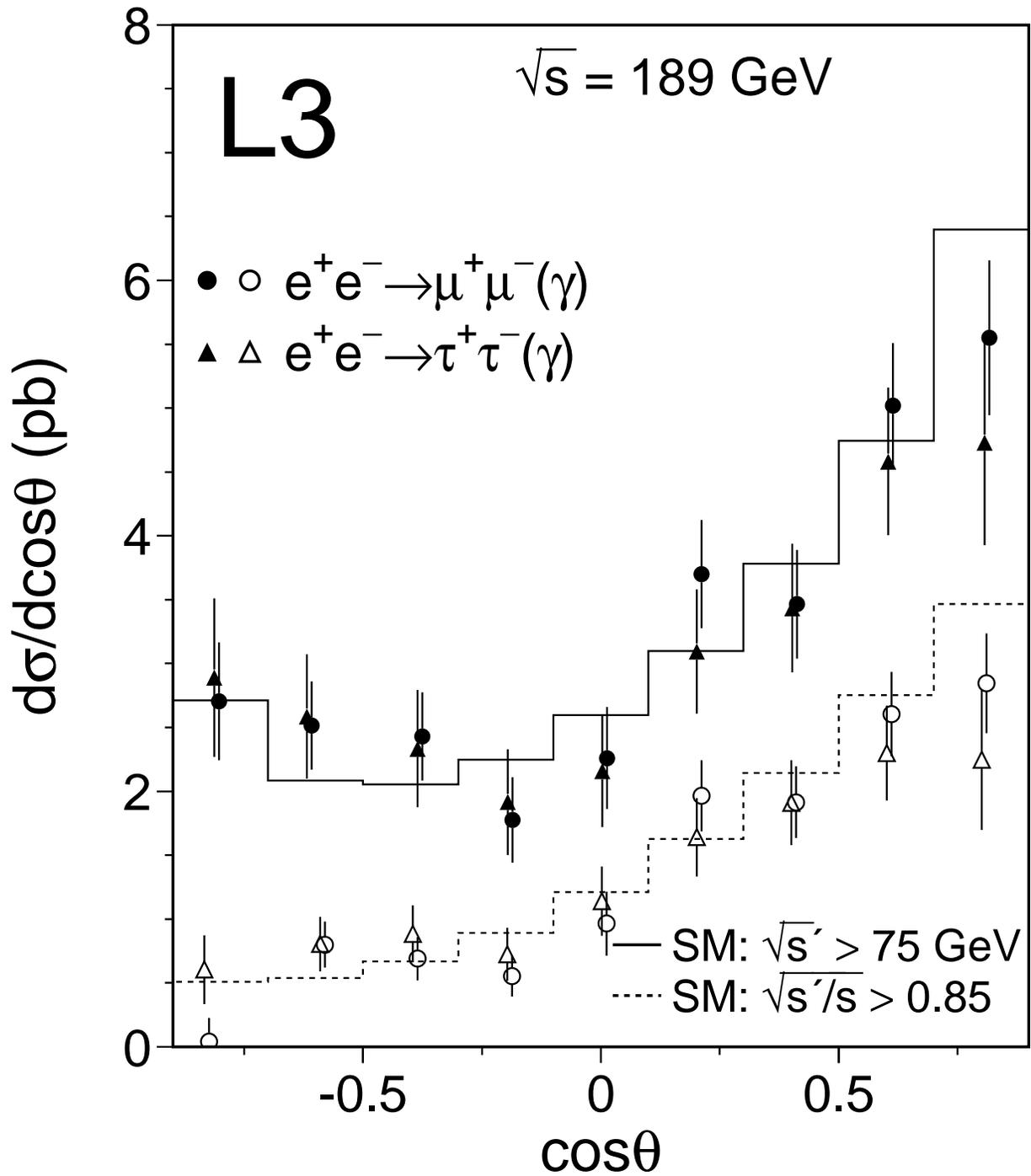}
    \parbox{\capwidth}{
      \caption[]{
        Differential cross section for the inclusive (solid symbols) and
        high-energy (open symbols) event samples,
        for the processes {$\epem\rightarrow\mumu(\gamma)$} and
        {$\epem\rightarrow\tautau(\gamma)$} at $\sqrt{s} = 189\GeV$.
        The lines indicate the \SM\ predictions.}
      \label{fig:ll_dsig}}
  \end{center}
\end{figure}

Table~\ref{tab:afb} summarises the numbers of forward and backward events,
the forward-backward asymmetry measurements, and their statistical and systematic
uncertainties. The main contributions to the systematic uncertainty are the
uncertainties on the backgrounds and on the momentum reconstruction.
Figure~\ref{fig:le_xsec_afb} shows the comparison of the corrected asymmetries
to the {\SM} prediction.
Table~\ref{tab:sigdiff} lists the differential cross sections at 183 \GeV\ and
189 \GeV, compared to their \SM\ predictions. The 189 \GeV\ distributions are
displayed in Figure~\ref{fig:ll_dsig}.

\subsection{\boldmath $\epem\rightarrow\tautau(\gamma)$}

\subsubsection*{Event selection}

Taus are identified as narrow, low multiplicity jets, containing at least one
charged particle. Tau jets are formed by matching the energy depositions in the
electromagnetic and hadron calorimeters with tracks in the central tracker and
the muon spectrometer.
Events containing two jets within the polar angular range $|\cos{\theta}|<0.92$
are accepted. The reconstruction of \sqrtsp\ follows the procedure
described in Section~\ref{sec:muon} using the polar angles of the two tau jets,
requiring at least 10 \GeV\ for observed photons.

Hadronic events are removed by requiring at most 16 calorimetric clusters with an
energy exceeding 100 \MeV\ each and at most 9 tracks in the central tracker.
Events containing two electrons or two muons are rejected.
Electrons are identified by a cluster in the electromagnetic calorimeter with an
energy greater than 2.5 \GeV\ and an electromagnetic shower shape, a matched track,
and less than 2.5 \GeV\ deposited in the hadron calorimeter. Muons are
identified by a track in the muon spectrometer and a minimum-ionising particle
signature in the calorimeters.
Bhabha events are further rejected by requiring the electromagnetic energy of the
highest-energy jet and the other jet to be less than 0.375 $\sqrtsp$ and 0.25
$\sqrtsp$, respectively.
In addition, the acoplanarity of the two jets must be larger than 0.2 degrees.

To reject background from two-photon collision processes the most energetic jet
must have an energy greater than 0.24 $E_{\mathrm{beam}}$. The distribution of
this quantity is shown in Figure~\ref{fig:sele}c for the data taken at 189 \GeV.
The energy of reconstructed muons is required to be less than 0.4 \sqrtsp.
To reject leptonic final states from W-pair production the acoplanarity of the two
tau jets must be less than 10 degrees.
Background from cosmic muons is reduced using both scintillation counter
information and the distance of the muon tracks from the beam axis.
Figure~\ref{fig:spri}c shows the reconstructed \sqrtsp\ distribution for the data
taken at 189 \GeV.

\subsubsection*{Cross section}

Selection efficiencies and background contaminations are listed in
Table~\ref{tab:sele}.
The numbers of selected events, as well as the cross sections and corresponding
statistical and systematic uncertainties for the different event
samples, are listed in Table~\ref{tab:xsec}. The systematic uncertainties
originate mainly from uncertainties in the event selection, in particular on the
rejection of Bhabha events.
Figure~\ref{fig:le_xsec_afb} shows the comparison to the {\SM} prediction.

\subsubsection*{Forward-backward asymmetry}

For the high-energy sample, the forward-backward asymmetry is determined using an
unbinned maximum-likelihood fit of Equation~\ref{eq:cost-afbfit} to events with
unambiguous charge assignment.
The background from other final states and from events with hard ISR photons is
corrected for in the fit procedure.
The fitted asymmetry is corrected for charge confusion.
For the inclusive sample, the forward-backward asymmetry is determined from the
differential cross section as described in Section~\ref{sec:mmafb}.
For both the inclusive and high-energy samples, the charge confusion per event
is estimated, from the data, to be less than 0.5\%.

Table~\ref{tab:afb} lists the number of forward and backward events and the
results of the asymmetry measurements.
Figure~\ref{fig:le_xsec_afb} shows the comparison of the measured asymmetries to
their \SM\ predictions.
Table~\ref{tab:sigdiff} lists the differential cross sections at 
183 \GeV\ and 189 \GeV, compared to their \SM\
predictions. The 189 \GeV\ distributions are displayed in Figure~\ref{fig:ll_dsig}.

\subsection{\boldmath $\epem\rightarrow\epem(\gamma)$}

\subsubsection*{Event selection}

Electron candidates are recognised by an energy deposition in the
electromagnetic calorimeter with at least 15 associated hits in the central
tracking chamber within a three degree azimuthal angular range.

Bhabha events are selected by requiring the two highest energy electron candidates
to be contained in the polar angular range $44^\circ < \theta < 136^\circ$, and to
have an energy greater than $0.5 \, E_{\mathrm{beam}}$ and 15~{\GeV}, respectively.
Figure~\ref{fig:sele}d shows the energy of the highest energy electron candidate,
normalised to the beam energy for events selected at 189~{\GeV}.

The acollinearity angle is calculated from the directions of the two electrons. Its
distribution is shown in Figure~\ref{fig:spri}d for events selected at
189 \GeV.

\subsubsection*{Cross section}

\begin{figure}[p]
  \begin{center}
    \includegraphics[height=0.45\textheight]{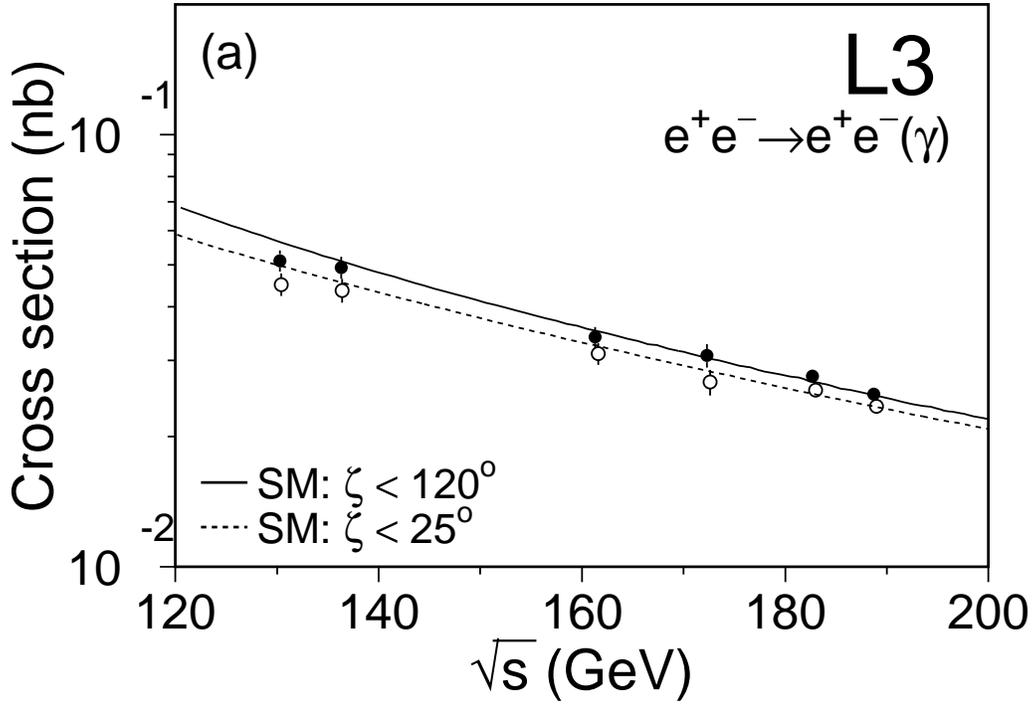}
    \includegraphics[height=0.45\textheight]{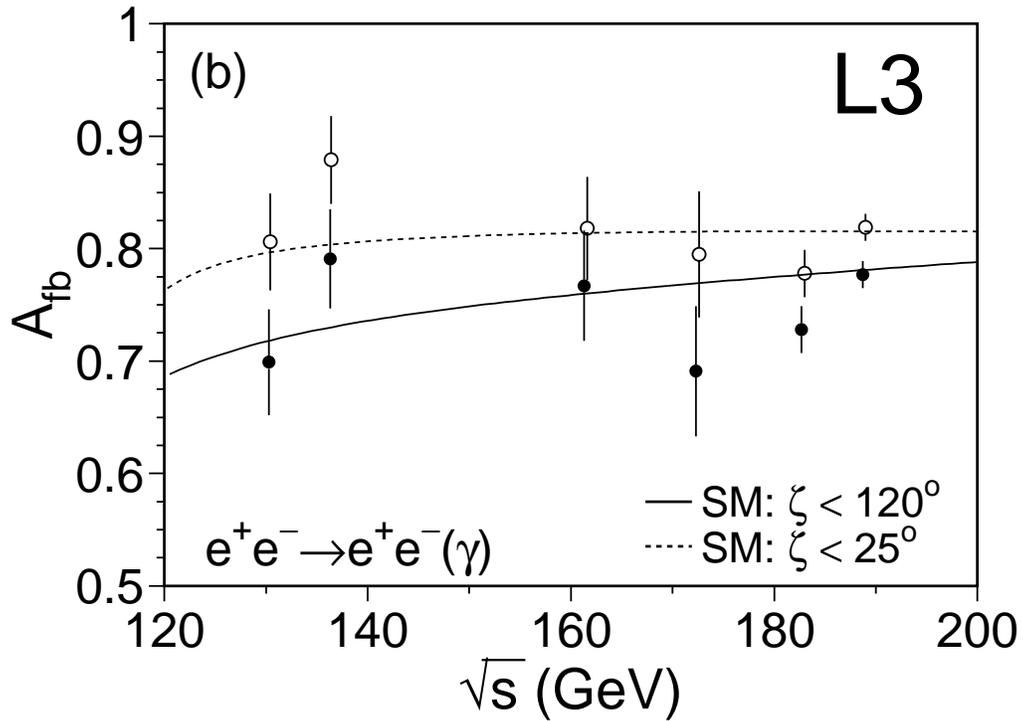}
    \parbox{\capwidth}{
      \caption[]{
        Cross sections (a) and forward-backward asymmetries (b) of the process
        {$\epem\rightarrow\epem(\gamma)$} 
        for the inclusive (solid symbols) and the high-energy sample (open symbols).
        The {\SM} predictions are shown as a solid line for the inclusive sample and
        as a dashed line for the high-energy sample. The two electrons are
        required to be inside $44^\circ < \theta < 136^\circ$.}
      \label{fig:ee_xsec_afb}}
  \end{center}
\end{figure}

The selection efficiencies within the fiducial volume and the background
contributions are listed in Table~\ref{tab:sele}. The background is dominated by
tau-pair production. Table~\ref{tab:xsec} lists the numbers of selected events,
and the measured cross sections with their statistical and systematic
uncertainties, for the various centre-of-mass energies. The systematic
errors are dominated by uncertainties on the event selection. The cross sections
are compared to their \SM\ prediction in Figure~\ref{fig:ee_xsec_afb}.

\subsubsection*{Forward-backward asymmetry}

The forward-backward asymmetry is extracted from the differential cross section.
The selection criteria for electron candidates are tightened to improve the charge
determination. 
The electron direction is obtained from both tracks in the event.
The charge confusion probability is determined~\cite{l3-69} from the data to be
$(2.8\pm0.3)$\% for the 130--136 \GeV\ data, $(4.3\pm0.3)$\% for the 183 \GeV\
data, and $(5.1\pm0.2)$\% for the 189 \GeV\ data, and is corrected for in the
determination of the differential cross section. The validity of the method has
been verified using a sample of dimuon events collected at the Z pole in 1995,
for which the charge is measured precisely in the muon spectrometer.

\begin{figure}[p]
  \begin{center}
    \includegraphics[width=1.\textwidth]{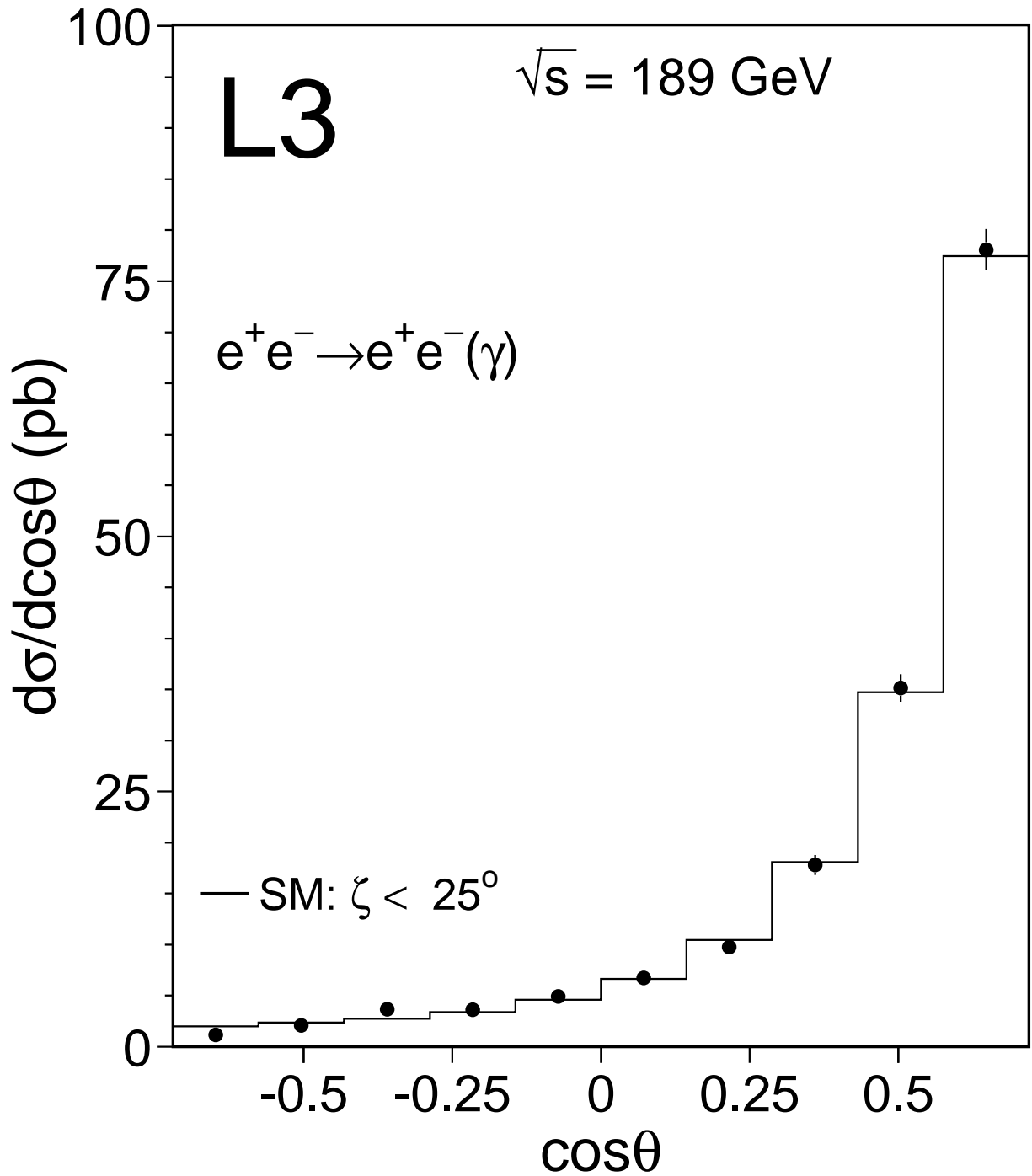}
    \parbox{\capwidth}{
      \caption[]{
        Differential cross section for the high-energy event sample, for the
        process {$\epem\rightarrow\epem(\gamma)$} at $\sqrt{s} = 
        189 \GeV$. The line indicates the \SM\ prediction.}
      \label{fig:ee_dsig}}
  \end{center}
\end{figure}

Table~\ref{tab:afb} summarises the numbers of forward and backward events and the
asymmetry measurements. The systematic error on the asymmetry measurements is
dominated by the uncertainty on the charge confusion.
Figure~\ref{fig:ee_xsec_afb} shows the comparison of the measured asymmetries to
their {\SM} predictions.
Table~\ref{tab:sigdiffe} lists the differential cross sections for the
high-energy samples at 183 \GeV\ and 189 \GeV, compared to their \SM\
predictions. The 189 \GeV\ distribution is displayed in
Figure~\ref{fig:ee_dsig}.

\section{Summary and Conclusion}

Based on an integrated luminosity of 243.7~\pb\ collected at {\CoM} energies
between 130.0~{\GeV} and 188.7~{\GeV}, we select 25864 hadronic and
8573 lepton-pair
events. The data are used to measure cross sections and leptonic
forward-backward asymmetries. The measurements are performed for the inclusive
event sample and for the high-energy sample. The results are in good agreement
with {\SM} predictions.

\section*{Acknowledgements}

We wish to congratulate the CERN accelerator divisions for the successful
upgrade of the LEP machine and to express our gratitude for the excellent
performance of the machine. We acknowledge the effort of the engineers,
technicians and support staff who have participated in the construction and
maintenance of this experiment.


\clearpage
\typeout{   }     
\typeout{Using author list for paper 195 -?}
\typeout{$Modified: Tue Nov 23 09:52:18 1999 by clare $}
\typeout{!!!!  This should only be used with document option a4p!!!!}
\typeout{   }
%
%
%
%
%
%

\newcount\tutecount  \tutecount=0
\def\tutenum#1{\global\advance\tutecount by 1 \xdef#1{\the\tutecount}}
\def\tute#1{$^{#1}$}
\tutenum\aachen            
\tutenum\nikhef            
\tutenum\mich              
\tutenum\lapp              
\tutenum\basel             
\tutenum\lsu               
\tutenum\beijing           
\tutenum\berlin            
\tutenum\bologna           
\tutenum\tata              
\tutenum\ne                
\tutenum\bucharest         
\tutenum\budapest          
\tutenum\mit               
\tutenum\debrecen          
\tutenum\florence          
\tutenum\cern              
\tutenum\wl                
\tutenum\geneva            
\tutenum\hefei             
\tutenum\seft              
\tutenum\lausanne          
\tutenum\lecce             
\tutenum\lyon              
\tutenum\madrid            
\tutenum\milan             
\tutenum\moscow            
\tutenum\naples            
\tutenum\cyprus            
\tutenum\nymegen           
\tutenum\caltech           
\tutenum\perugia           
\tutenum\cmu               
\tutenum\prince            
\tutenum\rome              
\tutenum\peters            
\tutenum\salerno           
\tutenum\ucsd              
\tutenum\santiago          
\tutenum\sofia             
\tutenum\korea             
\tutenum\alabama           
\tutenum\utrecht           
\tutenum\purdue            
\tutenum\psinst            
\tutenum\zeuthen           
\tutenum\eth               
\tutenum\hamburg           
\tutenum\taiwan            
\tutenum\tsinghua          
{
\parskip=0pt
\noindent
{\bf The L3 Collaboration:}
\ifx\selectfont\undefined
 \baselineskip=10.8pt
 \baselineskip\baselinestretch\baselineskip
 \normalbaselineskip\baselineskip
 \ixpt
\else
 \fontsize{9}{10.8pt}\selectfont
\fi
\medskip
\tolerance=10000
\hbadness=5000
\raggedright
\hsize=162truemm\hoffset=0mm
\def\r{\rlap,}
\noindent

M.Acciarri\r\tute\milan\
P.Achard\r\tute\geneva\ 
O.Adriani\r\tute{\florence}\ 
M.Aguilar-Benitez\r\tute\madrid\ 
J.Alcaraz\r\tute\madrid\ 
G.Alemanni\r\tute\lausanne\
J.Allaby\r\tute\cern\
A.Aloisio\r\tute\naples\ 
M.G.Alviggi\r\tute\naples\
G.Ambrosi\r\tute\geneva\
% ...
\newpage
\begin{list}{A}{\itemsep=0pt plus 0pt minus 0pt\parsep=0pt plus 0pt minus 0pt
                \topsep=0pt plus 0pt minus 0pt}
\item[\aachen]
 I. Physikalisches Institut, RWTH, D-52056 Aachen, FRG$^{\S}$\\
 III. Physikalisches Institut, RWTH, D-52056 Aachen, FRG$^{\S}$
\item[\nikhef] National Institute for High Energy Physics, NIKHEF, 
     and University of Amsterdam, NL-1009 DB Amsterdam, The Netherlands
\item[\mich] University of Michigan, Ann Arbor, MI 48109, USA
\item[\lapp] Laboratoire d'Annecy-le-Vieux de Physique des Particules, 
     LAPP,IN2P3-CNRS, BP 110, F-74941 Annecy-le-Vieux CEDEX, France
\item[\basel] Institute of Physics, University of Basel, CH-4056 Basel,
     Switzerland
\item[\lsu] Louisiana State University, Baton Rouge, LA 70803, USA
\item[\beijing] Institute of High Energy Physics, IHEP, 
  100039 Beijing, China$^{\triangle}$ 
\item[\berlin] Humboldt University, D-10099 Berlin, FRG$^{\S}$
\item[\bologna] University of Bologna and INFN-Sezione di Bologna, 
     I-40126 Bologna, Italy
\item[\tata] Tata Institute of Fundamental Research, Bombay 400 005, India
\item[\ne] Northeastern University, Boston, MA 02115, USA
\item[\bucharest] Institute of Atomic Physics and University of Bucharest,
     R-76900 Bucharest, Romania
\item[\budapest] Central Research Institute for Physics of the 
     Hungarian Academy of Sciences, H-1525 Budapest 114, Hungary$^{\ddag}$
\item[\mit] Massachusetts Institute of Technology, Cambridge, MA 02139, USA
\item[\debrecen] KLTE-ATOMKI, H-4010 Debrecen, Hungary$^\P$
\item[\florence] INFN Sezione di Firenze and University of Florence, 
     I-50125 Florence, Italy
\item[\cern] European Laboratory for Particle Physics, CERN, 
     CH-1211 Geneva 23, Switzerland
\item[\wl] World Laboratory, FBLJA  Project, CH-1211 Geneva 23, Switzerland
\item[\geneva] University of Geneva, CH-1211 Geneva 4, Switzerland
\item[\hefei] Chinese University of Science and Technology, USTC,
      Hefei, Anhui 230 029, China$^{\triangle}$
\item[\seft] SEFT, Research Institute for High Energy Physics, P.O. Box 9,
      SF-00014 Helsinki, Finland
\item[\lausanne] University of Lausanne, CH-1015 Lausanne, Switzerland
\item[\lecce] INFN-Sezione di Lecce and Universit\'a Degli Studi di Lecce,
     I-73100 Lecce, Italy
\item[\lyon] Institut de Physique Nucl\'eaire de Lyon, 
     IN2P3-CNRS,Universit\'e Claude Bernard, 
     F-69622 Villeurbanne, France
\item[\madrid] Centro de Investigaciones Energ{\'e}ticas, 
     Medioambientales y Tecnolog{\'\i}cas, CIEMAT, E-28040 Madrid,
     Spain${\flat}$ 
\item[\milan] INFN-Sezione di Milano, I-20133 Milan, Italy
\item[\moscow] Institute of Theoretical and Experimental Physics, ITEP, 
     Moscow, Russia
\item[\naples] INFN-Sezione di Napoli and University of Naples, 
     I-80125 Naples, Italy
\item[\cyprus] Department of Natural Sciences, University of Cyprus,
     Nicosia, Cyprus
\item[\nymegen] University of Nijmegen and NIKHEF, 
     NL-6525 ED Nijmegen, The Netherlands
\item[\caltech] California Institute of Technology, Pasadena, CA 91125, USA
\item[\perugia] INFN-Sezione di Perugia and Universit\'a Degli 
     Studi di Perugia, I-06100 Perugia, Italy   
\item[\cmu] Carnegie Mellon University, Pittsburgh, PA 15213, USA
\item[\prince] Princeton University, Princeton, NJ 08544, USA
\item[\rome] INFN-Sezione di Roma and University of Rome, ``La Sapienza",
     I-00185 Rome, Italy
\item[\peters] Nuclear Physics Institute, St. Petersburg, Russia
\item[\salerno] University and INFN, Salerno, I-84100 Salerno, Italy
\item[\ucsd] University of California, San Diego, CA 92093, USA
\item[\santiago] Dept. de Fisica de Particulas Elementales, Univ. de Santiago,
     E-15706 Santiago de Compostela, Spain
\item[\sofia] Bulgarian Academy of Sciences, Central Lab.~of 
     Mechatronics and Instrumentation, BU-1113 Sofia, Bulgaria
\item[\korea] Center for High Energy Physics, Adv.~Inst.~of Sciences
     and Technology, 305-701 Taejon,~Republic~of~{Korea}
\item[\alabama] University of Alabama, Tuscaloosa, AL 35486, USA
\item[\utrecht] Utrecht University and NIKHEF, NL-3584 CB Utrecht, 
     The Netherlands
\item[\purdue] Purdue University, West Lafayette, IN 47907, USA
\item[\psinst] Paul Scherrer Institut, PSI, CH-5232 Villigen, Switzerland
\item[\zeuthen] DESY, D-15738 Zeuthen, 
     FRG
\item[\eth] Eidgen\"ossische Technische Hochschule, ETH Z\"urich,
     CH-8093 Z\"urich, Switzerland
\item[\hamburg] University of Hamburg, D-22761 Hamburg, FRG
\item[\taiwan] National Central University, Chung-Li, Taiwan, China
\item[\tsinghua] Department of Physics, National Tsing Hua University,
      Taiwan, China
\item[\S]  Supported by the German Bundesministerium 
        f\"ur Bildung, Wissenschaft, Forschung und Technologie
\item[\ddag] Supported by the Hungarian OTKA fund under contract
numbers T019181, F023259 and T024011.
\item[\P] Also supported by the Hungarian OTKA fund under contract
  numbers T22238 and T026178.
\item[$\flat$] Supported also by the Comisi\'on Interministerial de Ciencia y 
        Tecnolog{\'\i}a.
\item[$\sharp$] Also supported by CONICET and Universidad Nacional de La Plata,
        CC 67, 1900 La Plata, Argentina.
\item[$\diamondsuit$] Also supported by Panjab University, Chandigarh-160014, 
        India.
\item[$\triangle$] Supported by the National Natural Science
  Foundation of China.
\item[\dag] Deceased.
\end{list}
}
\vfill






\clearpage

\begin{table}[p] 
  \begin{center}
    \renewcommand{\arraystretch}{1.0}
    \begin{tabular}{|l|r|r|r|r|r|r|r|r|}
      \hline
      & \multicolumn{4}{c|}{inclusive (\%)} & \multicolumn{4}{c|}{high energy (\%)} \\
      \multicolumn{1}{|c|}{$\sqrt{s}$ (\GeV)}
      & \multicolumn{1}{c}{130.0} & \multicolumn{1}{c}{136.1}
      & \multicolumn{1}{c}{182.7} & 188.7
      & \multicolumn{1}{c}{130.0} & \multicolumn{1}{c}{136.1}
      & \multicolumn{1}{c}{182.7} & 188.7 \\
\hline
\hline
\multicolumn{9}{|c|}{\boldmath $\epem\rightarrow\mbox{\textbf{hadrons}}(\gamma)$} \\
\hline
 Selection Efficiency  & 97.4 & 97.2 & 90.0 & 89.2 & 93.7 & 93.8 & 88.2 & 88.1 \\
 Two Photon Background &  1.8 &  1.8 &  2.5 &  2.8 &  1.4 &  1.8 &  1.8 &  1.9 \\
 $\Wp\Wm$ Background   &  --- &  --- &  4.4 &  5.2 &  --- &  --- &  6.9 &  8.1 \\
 Other Background      &  0.2 &  0.2 &  1.3 &  1.4 &  0.3 &  0.4 &  0.9 &  1.1 \\
 ISR Contamination     &  0.2 &  0.1 &  0.2 &  0.2 & 17.7 & 17.0 & 11.4 & 11.2 \\
\hline
\hline
\multicolumn{9}{|c|}{\boldmath $\epem\rightarrow\mumu(\gamma)$} \\
\hline
 Selection Efficiency  & 68.6 & 65.9 & 47.4 & 61.4 & 78.8 & 73.8 & 63.9 & 75.4 \\
 Two Photon Background &  2.0 &  2.8 &  4.8 & 10.0 &  0.8 &  1.5 &  1.8 &  2.9 \\
 $\Wp\Wm$ Background   &  --- &  --- &  0.7 &  2.9 &  --- &  --- &  0.5 &  2.5 \\
 Cosmic Background     &  0.9 &  1.1 &  2.0 &  0.5 &  0.9 &  2.4 &  1.8 &  0.4 \\
 Other Background      &  0.4 &  0.7 &  2.8 &  2.8 &  0.2 &  0.1 &  1.2 &  1.3 \\
 ISR Contamination     &  0.4 &  0.4 &  0.4 &  0.3 &  8.0 &  6.7 &  4.1 &  4.2 \\
\hline
\hline
\multicolumn{9}{|c|}{\boldmath $\epem\rightarrow\tautau(\gamma)$} \\
\hline
 Selection Efficiency  & 45.9 & 38.7 & 35.2 & 34.8 & 51.0 & 45.0 & 47.2 & 47.3 \\
 Two Photon Background &  2.3 &  2.1 &  7.2 & 7.4  &  1.3 &  1.2 &  2.6 &  2.2 \\
 Other Background      &  3.0 &  5.5 &  4.8 & 7.2  &  2.2 &  3.7 &  3.7 &  5.8 \\
 ISR Contamination     &  0.5 &  0.5 &  0.5 & 0.4  &  7.6 &  7.3 &  5.2 &  5.2 \\
\hline
\hline
\multicolumn{9}{|c|}{\boldmath $\epem\rightarrow\epem(\gamma)$} \\
\hline
 Selection Efficiency  & 97.9 & 97.6 & 96.4 & 97.5 & 98.0 & 97.6 & 95.9 & 97.1 \\
 $\tautau$ Background  &  1.3 &  1.5 &  1.3 &  1.3 &  1.1 &  1.4 &  1.3 &  1.3 \\
 Other Background      &  0.6 &  1.1 &  1.2 &  1.3 &  0.6 &  0.5 &  0.7 &  0.6 \\
\hline
    \end{tabular} \vskip 1.cm
    \parbox{\capwidth}{
      \caption[]{
        Selection efficiencies and background fractions for the inclusive
        and the high-energy
        event samples of the reactions {$\epem\rightarrow\mbox{hadrons}(\gamma)$},
        {$\epem\rightarrow\mumu(\gamma)$},
        {$\epem\rightarrow\tautau(\gamma)$} and
        {$\epem\rightarrow\epem(\gamma)$}. For Bhabha scattering the selection
        efficiencies are given for $44^\circ < \theta < 136^\circ$.}
      \label{tab:sele}}
  \end{center}
\end{table}

\clearpage
\begin{table}[p] 
  \begin{center}
    \begin{tabular}{|c|c|c|r@{$\,\pm\,$}l|c|c|r@{$\,\pm\,$}l|c|}
\hline
\multicolumn{2}{|c|}{} & \multicolumn{4}{c|}{inclusive}
                       & \multicolumn{4}{c|}{high energy} \\ 
  \multicolumn{1}{|c}{$\sqrt{s}$~(\GeV)} & \multicolumn{1}{c|}{$\mathcal{L}$~(\pb)} &
  \multicolumn{1}{c}{\Nsel} & \multicolumn{2}{c}{$\sigma$~(pb)} &
  \multicolumn{1}{c|}{\sigsm~(pb)} &
  \multicolumn{1}{c}{\Nsel} & \multicolumn{2}{c}{$\sigma$~(pb)} &
  \multicolumn{1}{c|}{\sigsm~(pb)} \\
\hline
\hline
\multicolumn{10}{|c|}{\boldmath $\epem\rightarrow\mbox{\textbf{hadrons}}(\gamma)$} \\
\hline
 130.0        &   6.1        &   1972 &     326.0 & 7.5$\pm 1.9$ &    329.5
                             &    632 &      84.2 & 4.4$\pm 1.0$ &     83.5  \\
 136.1        &   5.8        &   1571 &     274.4 & 7.0$\pm 1.8$ &    272.0
                             &    460 &      66.6 & 3.9$\pm 0.8$ &     66.9  \\
 161.3        &  10.0        &   1542 &     152.5 & 4.1$\pm 1.7$ &    151.8
                             &    423 &      37.3 & 2.2$\pm 0.7$ &     35.4  \\
 172.3        &   8.5        &   1064 &     121.2 & 4.1$\pm 1.3$ &    124.5
                             &    248 &      28.2 & 2.2$\pm 0.6$ &     28.8  \\
 182.7        &  54.9        &   5626 &     105.2 & 1.5$\pm 0.5$ &    105.7
                             &   1505 &      24.7 & 0.8$\pm 0.4$ &     24.3  \\
 188.7        & 173.4        &  16695 &      98.2 & 0.8$\pm 0.4$ &     96.9
                             &   4517 &      23.1 & 0.4$\pm 0.3$ &     22.2  \\
\hline
\hline
\multicolumn{10}{|c|}{\boldmath $\epem\rightarrow\mumu(\gamma)$} \\
\hline
 130.1        &  6.1         &     91 &    21.0 & 2.3$\pm 1.0 $  &  20.9
                             &     44 &     8.2 & 1.4$\pm 0.2 $  &  8.5   \\
 136.1        &  5.9         &     70 &    17.5 & 2.2$\pm 0.9 $  &  17.8
                             &     33 &     6.9 & 1.4$\pm 0.3 $  &  7.3   \\
 161.3        & 10.9         &     94 &    12.5 & 1.4$\pm 0.5 $  &  10.9
                             &     41 &    4.59 & 0.84$\pm 0.18$ &  4.70  \\
 172.1        & 10.2         &     67 &     9.2 & 1.3$\pm 0.4 $  &  9.2
                             &     32 &    3.60 & 0.75$\pm 0.14$ &  4.00  \\
 182.7        & 50.5         &    197 &    7.34 & 0.59$\pm 0.27$ &  7.90
                             &    111 &    3.09 & 0.33$\pm 0.14$ &  3.47  \\
 188.7        & 167.4        &    893 &    7.28 & 0.29$\pm 0.19$ &  7.29
                             &    420 &    2.92 & 0.16$\pm 0.06$ &  3.22  \\
\hline
\hline
\multicolumn{10}{|c|}{\boldmath $\epem\rightarrow\tautau(\gamma)$} \\
\hline
 130.1        &   6.1        &     66 &    22.1  & 2.9$\pm 0.5 $  &  20.9
                             &     35 &     9.8  & 1.9$\pm 0.3 $  &   8.5  \\
 136.1        &   5.9        &     43 &    17.1  & 2.8$\pm 0.5 $  &  17.8
                             &     23 &     7.5  & 1.8$\pm 0.3 $  &   7.3  \\
 161.3        &   9.8        &     45 &    10.4  & 2.0$\pm 0.7 $  &  10.9
                             &     25 &     4.6  & 1.1$\pm 0.3 $  &   4.7  \\
 172.1        &   9.7        &     45 &    11.0  & 2.0$\pm 0.8 $  &   9.2
                             &     23 &     4.3  & 1.1$\pm 0.3 $  &   4.0  \\
 182.7        &  55.5        &    174 &     7.77 & 0.68$\pm 0.17$ &   7.89
                             &    108 &     3.62 & 0.40$\pm 0.06$ &   3.47 \\
 188.7        & 176.8        &    527 &     7.27 & 0.37$\pm 0.17$ &   7.28
                             &    309 &     3.18 & 0.21$\pm 0.07$ &   3.22 \\
\hline
\hline
\multicolumn{10}{|c|}{\boldmath $\epem\rightarrow\epem(\gamma)$}\\
\hline
 130.1        &   6.1        &    312 &   51.1 & 2.9$\pm 0.2$ &   56.5 
                             &    274 &   45.0 & 2.7$\pm 0.2$ &   49.7  \\
 136.1        &   5.8        &    281 &   49.3 & 2.9$\pm 0.2$ &   50.9 
                             &    248 &   43.6 & 2.8$\pm 0.2$ &   45.4  \\
 161.3        &  10.2        &    337 &   34.0 & 1.9$\pm 1.0$ &   35.1
                             &    289 &   31.1 & 1.8$\pm 0.9$ &   32.4  \\
 172.3        &   8.8        &    256 &   30.8 & 1.9$\pm 0.9$ &   30.3
                             &    207 &   26.7 & 1.8$\pm 0.8$ &   28.3  \\
 182.7        &  55.3        &   1506 &   27.6 & 0.7$\pm 0.2$ &   26.7 
                             &   1385 &   25.6 & 0.7$\pm 0.1$ &   25.0  \\
 188.7        & 175.9        &   4413 &   25.1 & 0.4$\pm 0.1$ &   24.9 
                             &   4097 &   23.5 & 0.4$\pm 0.1$ &   23.4  \\
\hline
    \end{tabular}
    \vskip 1.cm
    \parbox{\capwidth}{
      \caption[]{
        Number of selected events, {\Nsel}, measured cross sections, $\sigma$,
        statistical errors and systematic errors and the {\SM} predictions,
        {\sigsm}, of the reactions {$\epem\rightarrow\mbox{hadrons}(\gamma)$},
        {$\epem\rightarrow\mumu(\gamma)$},
        {$\epem\rightarrow\tautau(\gamma)$} and
        {$\epem\rightarrow\epem(\gamma)$}, for the inclusive
        and the high-energy event samples.
        The systematic errors do not include the uncertainty on the luminosity
        measurement.
        In the case of Bhabha scattering, both leptons have to be inside
        $44^{\circ} < \theta < 136^{\circ}$.
        The results for the 161--172 \GeV\ data have been taken from
        Reference~\cite{l3-117} and corrected using ZFITTER ($s$-channel
        processes) and BHAGENE (Bhabha scattering) to correspond to the
        kinematic cuts described in the text.}
      \label{tab:xsec}}
  \end{center}
\end{table}

\clearpage
\begin{table}[p] 
  \begin{center}
    \begin{tabular}{|c|cc|c|c|cc|c|c|}
      \hline
      & \multicolumn{4}{|c|}{inclusive}
      & \multicolumn{4}{|c|}{high energy} \\
      $\sqrt{s}$~(\GeV) & \Nf & \multicolumn{1}{c}{\Nb} & \multicolumn{1}{c}{\Afb} &
      \AfbSM
      & \Nf & \multicolumn{1}{c}{\Nb} & \multicolumn{1}{c}{\Afb} & \AfbSM \\
\hline
\hline
\multicolumn{9}{|c|}{\boldmath $\epem\rightarrow\mumu(\gamma)$} \\
\hline
           130.0 &  61 &  29 & $0.46\pm   0.13       \pm 0.03$ & 0.324
                 &  38 &   5 & $0.67\plmi{0.08}{0.11}\pm 0.02$ & 0.707\\
           135.9 &  47 &  22 & $0.44\pm   0.15       \pm 0.04$ & 0.324
                 &  29 &   4 & $0.75\plmi{0.06}{0.11}\pm 0.05$ & 0.686\\
           161.3 &  35 &  21 & $0.32\plmi{0.12}{0.14}\pm 0.05$ & 0.325
                 &  22 &   8 & $0.59\plmi{0.13}{0.17}\pm 0.05$ & 0.619\\
           172.1 &  23 &  16 & $0.19\plmi{0.16}{0.17}\pm 0.05$ & 0.319
                 &  15 &   9 & $0.31\plmi{0.18}{0.21}\pm 0.05$ & 0.598\\
           182.7 & 133 &  59 & $0.45\pm   0.10       \pm 0.04$ & 0.313
                 &  86 &  23 & $0.62\plmi{0.07}{0.09}\pm 0.02$ & 0.582\\
           188.7 & 537 & 320 & $0.30\pm   0.04       \pm 0.02$ & 0.310
                 & 312 &  91 & $0.58\pm   0.04       \pm 0.02$ & 0.573\\
\hline
\hline
\multicolumn{9}{|c|}{\boldmath $\epem\rightarrow\tautau(\gamma)$} \\
\hline                 
           130.1 &  36 &  17 & $0.42\pm   0.17       \pm 0.03$ & 0.324
                 &  22 &   3 & $0.78\plmi{0.10}{0.16}\pm 0.02$ & 0.707\\
           136.1 &  26 &  10 & $0.53\pm   0.22       \pm 0.01$ & 0.325
                 &  18 &   2 & $0.96\plmi{0.10}{0.17}\pm 0.03$ & 0.686\\
           161.3 &  15 &  12 & $0.19\plmi{0.19}{0.20}\pm 0.10$ & 0.325
                 &  12 &   4 & $0.97\plmi{0.03}{0.23}\pm 0.10$ & 0.619\\
           172.1 &  16 &  16 & $0.10\plmi{0.18}{0.19}\pm 0.10$ & 0.319
                 &   9 &   7 & $0.18\plmi{0.25}{0.27}\pm 0.10$ & 0.598\\
           182.7 &  96 &  52 & $0.26\pm   0.10       \pm 0.01$ & 0.313
                 &  62 &  29 & $0.53\plmi{0.10}{0.11}\pm 0.03$ & 0.582\\
           188.7 & 275 & 144 & $0.23\pm   0.06       \pm 0.02$ & 0.310
                 & 177 &  72 & $0.44\pm   0.06       \pm 0.02$ & 0.573\\
\hline
\hline
\multicolumn{9}{|c|}{\boldmath $\epem\rightarrow\epem(\gamma)$} \\
\hline
           130.0 & 214 &  44 & $0.699 \pm 0.047 \pm 0.005$ &  0.715
                 & 201 &  27 & $0.806 \pm 0.043 \pm 0.006$ &  0.799 \\
           136.1 & 199 &  29 & $0.791 \pm 0.044 \pm 0.005$ &  0.728
                 & 185 &  17 & $0.879 \pm 0.039 \pm 0.006$ &  0.804 \\
           161.3 & 240 &  43 & $0.767 \pm 0.049 \pm 0.012$ &  0.762
                 & 206 &  29 & $0.818 \pm 0.046 \pm 0.012$ &  0.809 \\
           172.1 & 203 &  51 & $0.691 \pm 0.058 \pm 0.012$ &  0.771
                 & 176 &  31 & $0.795 \pm 0.056 \pm 0.012$ &  0.812 \\
           182.7 &1020 & 207 & $0.728 \pm 0.021 \pm 0.004$ &  0.773
                 & 972 & 162 & $0.778 \pm 0.021 \pm 0.004$ &  0.813 \\
           188.7 &3045 & 546 & $0.777 \pm 0.012 \pm 0.007$ &  0.779
                 &2912 & 443 & $0.819 \pm 0.012 \pm 0.003$ &  0.815 \\
\hline
    \end{tabular} \vskip 0.5cm
    \parbox{\capwidth}{
      \caption[]{ Number of forward, \Nf, and backward events, \Nb,
        forward-backward asymmetries, \Afb, statistical and systematic
        errors and the {\SM} predictions, {\AfbSM}, of the reactions
        {$\epem\rightarrow\mumu(\gamma)$},
        {$\epem\rightarrow\tautau(\gamma)$} and 
        {$\epem\rightarrow\epem(\gamma)$} for the inclusive
        and the high-energy event samples.
        In the case of Bhabha scattering, both leptons have to be inside
        $44^\circ < \theta < 136^\circ$.
        The results for the 161--172 \GeV\ data have been taken from
        Reference~\cite{l3-117} and corrected using ZFITTER ($s$-channel
        processes) and BHAGENE (Bhabha scattering) to correspond to the
        kinematic cuts described in the text.}
      \label{tab:afb}}
  \end{center}
\end{table}

\begin{table}[tbh]
  \begin{center}
    \begin{tabular}{|r@{,}l|r@{$\pm$}l|r@{$\pm$}l|c|r@{$\pm$}l|r@{$\pm$}l|c|}
      \hline
      \multicolumn{2}{|c|}{$\cos\theta$ range} & \multicolumn{5}{c|}{182.7 \GeV} &
      \multicolumn{5}{c|}{188.7 \GeV} \\
      \multicolumn{2}{|c|}{} & \multicolumn{2}{c}{$\mumu(\gamma)$} &
      \multicolumn{2}{c}{$\tautau(\gamma)$} & SM &
      \multicolumn{2}{c}{$\mumu(\gamma)$} &
      \multicolumn{2}{c}{$\tautau(\gamma)$} & SM \\
      \hline
      \hline
      \multicolumn{12}{|c|}{$\sqrtsp > 75 \GeV$}\\
      \hline
      {[}-0.90 & -0.70{]} & 0.339 & 0.172 & 0.491 & 0.218 & 0.584 & 0.541 & 0.093 & 0.578 & 0.126 & 0.542 \\
      {[}-0.70 & -0.50{]} & 0.261 & 0.101 & 0.390 & 0.157 & 0.452 & 0.503 & 0.070 & 0.517 & 0.099 & 0.417 \\
      {[}-0.50 & -0.30{]} & 0.508 & 0.131 & 0.577 & 0.178 & 0.446 & 0.486 & 0.070 & 0.467 & 0.093 & 0.411 \\
      {[}-0.30 & -0.10{]} & 0.242 & 0.112 & 0.753 & 0.199 & 0.490 & 0.355 & 0.070 & 0.383 & 0.084 & 0.450 \\
      {[}-0.10 &\phm0.10{]} & 0.867 & 0.228 & 0.559 & 0.166 & 0.567 & 0.452 & 0.081 & 0.432 & 0.088 & 0.519 \\
      {[} 0.10 &\phm0.30{]} & 0.443 & 0.143 & 0.461 & 0.155 & 0.677 & 0.740 & 0.087 & 0.619 & 0.098 & 0.620 \\
      {[} 0.30 &\phm0.50{]} & 1.315 & 0.215 & 0.743 & 0.186 & 0.825 & 0.693 & 0.087 & 0.687 & 0.102 & 0.757 \\
      {[} 0.50 &\phm0.70{]} & 0.769 & 0.161 & 0.853 & 0.200 & 1.035 & 1.004 & 0.101 & 0.916 & 0.117 & 0.950 \\
      {[} 0.70 &\phm0.90{]} & 1.005 & 0.289 & 1.868 & 0.393 & 1.391 & 1.110 & 0.124 & 0.946 & 0.165 & 1.280 \\
      \hline
      \hline
      \multicolumn{12}{|c|}{$\sqrtsp > 0.85 \sqrt{s}$}\\
      \hline
      {[}-0.90 & -0.70{]} & 0.115 & 0.101 & 0.103 & 0.098 & 0.105 & 0.008 & 0.037 & 0.121 & 0.055 & 0.102 \\
      {[}-0.70 & -0.50{]} & 0.057 & 0.044 & 0.092 & 0.065 & 0.113 & 0.160 & 0.036 & 0.161 & 0.043 & 0.108 \\
      {[}-0.50 & -0.30{]} & 0.139 & 0.066 & 0.209 & 0.087 & 0.142 & 0.138 & 0.034 & 0.177 & 0.045 & 0.134 \\
      {[}-0.30 & -0.10{]} & 0.080 & 0.049 & 0.328 & 0.108 & 0.191 & 0.111 & 0.032 & 0.145 & 0.042 & 0.179 \\
      {[}-0.10 &\phm0.10{]} & 0.396 & 0.127 & 0.272 & 0.101 & 0.262 & 0.193 & 0.050 & 0.228 & 0.055 & 0.243 \\
      {[} 0.10 &\phm0.30{]} & 0.190 & 0.078 & 0.255 & 0.099 & 0.353 & 0.393 & 0.056 & 0.328 & 0.061 & 0.326 \\
      {[} 0.30 &\phm0.50{]} & 0.703 & 0.139 & 0.375 & 0.119 & 0.464 & 0.383 & 0.056 & 0.382 & 0.067 & 0.429 \\
      {[} 0.50 &\phm0.70{]} & 0.427 & 0.111 & 0.529 & 0.143 & 0.597 & 0.521 & 0.066 & 0.460 & 0.075 & 0.551 \\
      {[} 0.70 &\phm0.90{]} & 0.415 & 0.164 & 1.079 & 0.299 & 0.752 & 0.569 & 0.079 & 0.450 & 0.113 & 0.694 \\
      \hline
    \end{tabular}
    \parbox{\capwidth}{
      \caption[]{Cross sections (in pb) for the processes
        $\epem\rightarrow\mumu(\gamma)$ and $\epem\rightarrow\tautau(\gamma)$ at
        183~\GeV\ and 189~\GeV\ in bins of $\cos\theta$, compared to their
        Standard Model predictions. Statistical and systematic uncertainties are
        combined.}
      \label{tab:sigdiff}}
  \end{center}
\end{table}
\begin{table}[bth]
  \begin{center}
    \begin{tabular}{|r@{,}l|r@{$\pm$}l|c|r@{$\pm$}l|c|}
      \hline
      \multicolumn{2}{|c|}{$\cos\theta$ range} & \multicolumn{2}{c}{182.7 \GeV}
      & SM & \multicolumn{2}{c}{188.7 \GeV} & SM \\
      \hline
      {[}-0.719& -0.575{]} &  0.404 & 0.157 &  0.306 &  0.167 & 0.075 &  0.290 \\
      {[}-0.575& -0.432{]} &  0.306 & 0.115 &  0.352 &  0.300 & 0.065 &  0.344 \\
      {[}-0.432& -0.288{]} &  0.532 & 0.127 &  0.433 &  0.526 & 0.070 &  0.395 \\
      {[}-0.288& -0.144{]} &  0.539 & 0.122 &  0.532 &  0.524 & 0.067 &  0.488 \\
      {[}-0.144&\phm0.000{]} &  0.930 & 0.153 &  0.712 &  0.708 & 0.075 &  0.664 \\
      {[} 0.000&\phm0.144{]} &  0.930 & 0.153 &  1.028 &  0.973 & 0.087 &  0.956 \\
      {[} 0.144&\phm0.288{]} &  1.638 & 0.199 &  1.609 &  1.405 & 0.103 &  1.509 \\
      {[} 0.288&\phm0.432{]} &  2.779 & 0.259 &  2.810 &  2.561 & 0.139 &  2.607 \\
      {[} 0.432&\phm0.575{]} &  5.025 & 0.347 &  5.350 &  5.054 & 0.194 &  5.003 \\
      {[} 0.575&\phm0.719{]} & 12.39  & 0.54  & 11.91  & 11.23  & 0.29  & 11.16  \\
      \hline
    \end{tabular}
    \parbox{\capwidth}{
      \caption[]{Cross sections (in pb) for $\zeta < 25^{\circ}$ for the process
        $\epem\rightarrow\epem(\gamma)$ at 183~\GeV\ and 189~\GeV\ in bins
        of $\cos\theta$, compared to their Standard Model
        predictions. Statistical and systematic uncertainties are combined.}
      \label{tab:sigdiffe}}
  \end{center}
\end{table}

\clearpage

\end{document}